\documentclass[a4paper,11pt]{article}
\usepackage{jheppub} 

\usepackage{amssymb,amsthm,amsmath}
\usepackage{textcomp}
\usepackage{color}      
\usepackage{slashed}    
\usepackage{verbatim}
\usepackage[normalem]{ulem} 
\usepackage{rotating}   
\usepackage{multirow}   
\usepackage{hyperref}
\usepackage{bm}
\begin{document} 
	
	\raggedbottom
	
	\title{Identify hadron anomalous couplings at colliders
	}

	\author[a]{   Chao-Qiang Geng
	}	
	\author[a,b,c]{   Chia-Wei Liu  
	}
	\author[a,d,e,f]{  Yue-Liang Wu
	}	
	\emailAdd{cqgeng@ucas.ac.cn, chiaweiliu@sjtu.edu.cn, ylwu@itp.ac.cn}
	
\affiliation[a]{School of Fundamental Physics and Mathematical Sciences, Hangzhou Institute for Advanced Study, UCAS, Hangzhou 310024, China}
 	\affiliation[b]{State Key Laboratory of Dark Matter Physics, 
	Tsung-Dao Lee Institute and School of Physics and Astronomy,
	Shanghai Jiao Tong University, Shanghai 200240, China} 
\affiliation[c]{Key Laboratory for Particle Astrophysics and Cosmology (MOE) and Shanghai Key Laboratory for Particle Physics and Cosmology, Tsung-Dao Lee Institute and School of Physics and Astronomy,
	Shanghai Jiao Tong University, Shanghai 200240, China}
\affiliation[d]{ Institute of Theoretical Physics, Chinese Academy of Sciences, Beijing 100190, China }
\affiliation[e]{International Centre for Theoretical Physics Asia-Pacific  (ICTP-AP), Beijing 100190, China}
\affiliation[f]{Taiji Laboratory for Gravitational Wave Universe (Beijing/Hangzhou), University of Chinese Academy of Sciences(UCAS), Beijing 100049, China }

	\date{\today}

	\abstract{
We investigate the identification of the Wess-Zumino-Witten (WZW) Lagrangian at colliders such as BESIII and the Super-$\tau$-Charm Facility. Our analysis concentrates on the radiative decays of $\eta$ and $\eta'$ mesons, including $\eta^{(\prime)} \to \gamma\gamma$, $\eta^{(\prime)} \to \gamma\ell^+ \ell^-$, $\eta^{(\prime)} \to \pi^+\pi^-\gamma$, and $\eta^{(\prime)} \to \pi^+\pi^-\ell^+\ell^-$, as well as semileptonic kaon decays such as $K^+ \to \pi^+\pi^- e^+ \nu_e$. Employing the hidden local symmetry framework to incorporate vector meson contributions, we compute the decay amplitudes and form factors.
For the decay $\eta \to \pi^+\pi^-\gamma$, the box anomaly dominates, and we find that the anomalous coupling can be experimentally determined to percent-level precision at BESIII. In contrast, vector meson contributions are significant in the decay $\eta' \to \pi^+\pi^-\gamma$. Using experimental data for $\eta' \to \pi^+\pi^-\gamma$, we obtain ${\cal B}_{\mathrm{box}}^{\text{exp}} = (1.70 \pm 0.05)\%$, which is approximately ten times larger than previously expected experimentally. We observe good agreement between our calculated anomalous couplings and experimental results. 
In kaon decays, WZW terms uniquely contribute to the form factor $H$, which can be extracted from parity-conserving decay distributions. While predictions at the chiral point closely match experimental values (e.g., $H^+ = -2.31$ versus $-2.27 \pm 0.10$), we find that intermediate vector meson states introduce substantial corrections, potentially as large as 25\%. We strongly advocate for revisiting these experiments to achieve improved precision in form factor extractions.
	}

	\maketitle

\section{Introduction}

The low-energy dynamic  of the strong interactions  is  effectively captured by chiral perturbation theory (\(\chi\)PT)~\cite{Weinberg:1978kz}, which incorporates the symmetries of the underlying Quantum Chromodynamics (QCD). A particularly rich component is the anomalous sector, arising from triangle and box anomalies, and described by the Wess--Zumino--Witten (WZW) Lagrangian in \(\chi\)PT~\cite{Wess:1971yu,Witten:1983tw}. In 
the 
Witten’s geometric formulation, this Lagrangian emerges from a five-dimensional Chern--Simons form~\cite{Wu:1986pr} whose boundary is the physical spacetime. It governs processes involving an odd number of pseudoscalar mesons, exemplified by decays such as \(\pi^0 \to \gamma \gamma\) and \(\eta \to \pi^+ \pi^- \gamma\)~\cite{Benayoun:1992ty,Petri:2010ea,Adler:1969gk}.

Significant opportunities for exploring anomalous processes are provided by BESIII and the future Super-$\tau$-Charm Facility (STCF)~\cite{Achasov:2023gey}. BESIII has already collected over a ten  billion $J/\psi$ events, enabling studies of $\eta^{(\prime)}$   through radiative decays with branching fractions  
$
\mathcal{B}(J/\psi \to \eta \gamma) = (1.090 \pm 0.013) \times 10^{-3}$ and $ 
\mathcal{B}(J/\psi \to \eta' \gamma) = (5.28 \pm 0.06) \times 10^{-3}
$~\cite{BESIII:2023fai}. 
The form factors and branching fractions of 
$\eta^{(\prime)} \to \gamma \ell^+ \ell^-$ and $\eta^{(\prime)} \to \pi^+ \pi^- \gamma$, with $\ell = e, \mu$, have been measured with good precision~\cite{BESIII:2015zpz,BESIII:2017kyd,BESIII:2020elh,BESIII:2021fos,BESIII:2024awu,BESIII:2025cky}. The STCF is expected to produce over a hundred times more $J/\psi$ events, significantly enhancing the sensitivity to rare decay modes by an order of magnitude~\cite{Achasov:2023gey}.

In this study, we investigate the feasibility of probing the WZW Lagrangian at current and forthcoming collider experiments. Unlike the pion case, the masses of the $\eta^{(\prime)}$ mesons are comparable to those of the light vector mesons, thus requiring the inclusion of vector meson corrections. Nevertheless, the low-energy predictions of $\chi$PT must be consistently recovered in the appropriate limit. The hidden local symmetry (HLS) framework offers a systematic and flexible methodology to incorporate these~\cite{Bando:1984ej}. This approach inherently contains the WZW structure while explicitly accounting for vector-meson degrees of freedom. In the low-energy regime, the original WZW Lagrangian emerges naturally from the equations of motion~\cite{Harada:2003jx}. Additionally, through appropriate parameter choices, the HLS framework can be simplified to well-known models, such as the vector meson dominance (VMD) approximation~\cite{Sakurai:1960ju}.
The HLS framework can be extended to encompass axial-vector mesons~\cite{Ma:2004sc} and  has notably been employed for extensive studies on hadronic vacuum polarization and isospin-breaking effects~\cite{BHLS}. Recently, consistent couplings to axions have also been integrated into the WZW Lagrangian, further broadening its scope of application~\cite{Bai:2024lpq}.

This paper is structured as follows. Section~\ref{sec2} outlines the formalism of the WZW Lagrangian and the HLS framework, emphasizing the simplest representative scenarios without quark mass effects. In Section~\ref{sec3}, we examine radiative $\eta$ meson decays driven by the WZW Lagrangian, highlighting the anticipated experimental sensitivities. Section~\ref{sec4} focuses on kaon weak decays induced by the same anomalous structure. Finally, our conclusions are presented in Section~\ref{sec5}.

	\section{Formalism}\label{sec2}
	
We start with the action of $\chi$PT. Its general action reads:
\begin{equation}\label{action}
	S = S_{\chi \text{PT} } + S_{\text{WZW}} .
\end{equation}
The leading term comes from the usual nonlinear $\sigma$ model:
\begin{equation}
	S _{\chi \text{PT} } = \int_{M^4}
	\frac{f_P^2}{8}  \text{Tr} \left( ( D_\mu U)^\dagger D^\mu U \right)  d^4 x
	+ \mathcal{O}(p^4) \,.
\end{equation}
The trace is over the $SU(3)$ flavor space of
\begin{equation}
	U = \exp\left( \frac{2i}{f_P} P \right)
	= \exp\left( \frac{2i}{f_P} \left[ \begin{array}{ccc}
		\frac{\eta_0}{\sqrt{3}} + \frac{\eta_8}{\sqrt{6}} + \frac{\pi^0}{\sqrt{2}} & \pi^{+} & K^{+} \\
		\pi^{-} & \frac{\eta_0}{\sqrt{3}} + \frac{\eta_8}{\sqrt{6}} - \frac{\pi^0}{\sqrt{2}} & K^0 \\
		K^{-} & \bar{K}^0 & \frac{\eta_0}{\sqrt{3}} - \frac{2 \eta_8}{\sqrt{6}}
	\end{array} \right] \right) ,
\end{equation}
and $D_\mu U = \partial_\mu U - i \mathcal{L}_ \mu U + i U \mathcal{R}_ \mu$, where ${\cal L}_ \mu$ and ${\cal R}_ \mu$ are gauge fields.
Here, $f_P$ is the meson decay constant.
By definition, $S_{\chi \text{PT} }$ contains all terms that are integrals over the usual spacetime $M^4$ and is gauge-invariant by itself. The second term, $S_{\text{WZW}}$, is the WZW action, living in the five-dimensional manifold with 
the 
gauge countered term in   $M^4$. 
	
	In parity-conserving processes, $S $ must be invariant under the parity transformation 
	\begin{equation}\label{1.4}
		(\vec{x}, P) \to (-\vec{x}, -P),
	\end{equation}
	where the minus sign for $P \to -P$ arises due to the intrinsic parity of the pseudoscalar mesons, known as the $(-1)^{N_P}$ transformation, with $N_P$ being the number of pseudoscalar mesons. As noted by Witten, $S_{\chi \text{PT} }$ is invariant under the $(-1)^{N_P}$ transformation 
	and $\vec{x} \to -\vec{x}$  separately, which is clearly not a symmetry shared by the QCD~\cite{Witten:1983tw}.

In general, even when terms with more derivatives are included, the $(-1)^{N_P}$ symmetry is still respected by $S_{\chi \text{PT} }$ due to the identity
\begin{equation}\label{N}
	\operatorname{Tr}\left( \prod_{n=1}^N ( \partial_{\mu_n} U) U^{-1} \right)
	= (-1)^N \operatorname{Tr}\left( \prod_{n=1}^N (\partial_{\mu_n} U^{-1}) U \right) .
\end{equation}
Here, $\partial_{\mu_n}$ represents the derivatives in $M^4$. To form a Lorentz-invariant term, all the $\mu_n$ must be contracted, and thus the total number $N$ must be even.
	 
To break the $(-1)^{N_P}$ symmetry, we need $N$ to be odd in eq.~\eqref{N}. In other words, a totally antisymmetric tensor  
with odd-dimension
is needed. Hence, $ S_{\text{WZW}}$ is introduced, which lives in the five-dimensional manifold, given by
\begin{equation}\label{WZW}
	S ^0_{\text{WZW}}
	\equiv 
	S_{\text{WZW}}
	\left(
	{\cal R} ={\cal L} = 0 
	\right)
	=   
\frac{N_c}{240 \pi^2}  \int _ {M^5} \text{Tr} \left(
	\alpha ^5 
	\right)
	\,,
\end{equation}
with  $N_c$ the  number of color and ${\cal L}, {\cal R}$ the gauge fields.     
The boundary of  $M^5$ is the usual four-dimensional spacetime $\partial M^5 = M^4$. The differential form $\alpha$ is defined by
\begin{equation}
	\alpha  = \frac{1}{i }d U U^{-1} \,. 
\end{equation}
By a straightforward computation, we have $d \, \text{Tr} \left(
\alpha ^5 
\right) = 0 $. Thus,   
$ \text{Tr} \left(
\alpha ^5 
\right)$ is locally exact according to the Poincaré lemma and the $M^5$ integral in eq.~\eqref{WZW} can be reduced to $
\partial M^5 = 
M^4$ according to the Stokes' theorem.
	
We note that there is  no gauge field  involved in the five-forms of $S^0_{\text{WZW}}$, and therefore it must be invariant under the parity transformation of eq.~\eqref{1.4}.
	The coordinate transformation $\vec{x} \to - \vec{x}$ would cause $S _{\text{WZW}}^0 \to - S_{\text{WZW}}^0$ due to the wedge product of forms. Using the cyclic property of the trace,   the $(-1)^{N_P}$ transformation introduces another minus sign, compensating the sign change from the former.

When gauge fields are considered, several terms need to be added to ensure that $S_{\text{WZW}}$ is gauge invariant, given by~\cite{Wu:1986pr,Witten:1983tw}
\begin{equation}
S_{\mathrm{WZW}}
\left(
U , {\cal R} ,{\cal L}
\right)
=
S_{\text{WZW}}^0
+
\frac{N_c}{48  \pi^2} 
\int _{M^4}    \operatorname{Tr} \left( {\cal I} \right)\,, 
\end{equation}
where
\begin{equation}\label{LQED}
\begin{aligned}
	{\cal I} = & \, i \left[ \frac{1}{2} \left( {\cal L} \alpha \right)^2 + {\cal L} \alpha {\cal L} U {\cal R} U^{-1} 
	+ \frac{1}{4} \left( {\cal L} U {\cal R} U^{-1} \right)^2  
 + \left( \mathcal{L}^3 - \alpha^2 {\cal L} \right) \left( \alpha + U \mathcal{R} U^{-1} \right) \right] \\
	& + d \mathcal{L} \alpha U \mathcal{R} U^{-1} 
	- (d \mathcal{L} \mathcal{L} + \mathcal{L} d \mathcal{L}) ( \alpha + U {\cal R} U^{-1} ) \\
	& + ( {\cal L} \leftrightarrow - {\cal R} , \alpha \to U^{-1} \alpha U , U \to U^{-1} , i \to -i ) \,,
\end{aligned}
\end{equation}
  ${\cal L} = {\cal L}_\mu d x^\mu$ and ${\cal R} = {\cal R}_\mu d x^\mu$. Here, the last term corresponds to the substitutions of the other terms by interchanging ${\cal L}$ and $-{\cal R}$, and so on.
Since
	four-forms    flip sign under $\vec{x} \to - \vec{x}$,  $S _{\text{WZW}}$ flips signs under the $(-1)^{N_P}$ transformation for parity-conserving processes. This can also be seen by  taking ${\cal L} = {\cal R}$  in eq.~\eqref{LQED}.

Parity-conserving processes with an odd number of pseudoscalar mesons would 
break the $(-1)^{N_P}$ symmetry and 
provide clear evidence of   $S_{\text{WZW}}$. On the other hand, for parity-violating processes, $(-1)^{N_P}$ may be broken by $S_{\chi \text{PT} }$ as well. To search for evidence in these channels, we must extract the form factors, as will be demonstrated.

In the standard model, we have the gauge fields of photons~$(A_\mu)$, the $Z$ boson~$(Z_\mu)$, and the $W$ boson~$(W^\pm_\mu)$, given by
\begin{equation}
\begin{aligned}
	\mathcal{L}_\mu &= e Q A_\mu + \frac{g_2}{\cos \theta_W} \left(T_z - \sin^2 \theta_W \right) Z_\mu + \frac{g_2}{\sqrt{2}} \left(W_\mu^{+} T_{+} + W_\mu^{-} T_{-}\right) \,, \\
	\mathcal{R}_\mu &= e Q A_\mu - \frac{g_2}{\cos \theta_W} \sin^2 \theta_W Z_\mu \,, 
\end{aligned}
\end{equation}
with $e^2= 4 \pi \alpha_{\text{em}}$ and $\alpha_{\text{em}}$ the fine structure constant.  
Here, $Q = \text{diag}\left( \frac{2}{3}, -\frac{1}{3}, -\frac{1}{3} \right)$, $T_z = \text{diag}\left( \frac{1}{2}, -\frac{1}{2}, -\frac{1}{2} \right)$, and
\begin{equation}
T_+ = T_-^\dagger = \left( \begin{array}{ccc}
	0 & V_{u d} & V_{u s} \\
	0 & 0 & 0 \\
	0 & 0 & 0
\end{array} \right) \,,
\end{equation}
with $V_{qq'}$ being the CKM matrix elements.

	The action in eq.~\eqref{action} is only valid at low energies. It provides an excellent approximation for pion decays. However,    the   masses of  $\eta$, $\eta'$ and kaons are comparable to those of the   vector  mesons, so the degrees of freedom of vector mesons need to be also considered. To estimate the effects of resonances, we adopt the HLS. The key idea of the HLS is to rewrite $U = \xi_L^\dagger \xi_R$. This generates a new type of local symmetry, such that the action is invariant under $\xi_{L,R} \to h(x) \xi_{L,R}$ with $h(x)^\dagger h (x)=1$. In this local symmetry, vector mesons $V$ play the role of  gauge bosons.
	The relevant Lagrangian  lives in $M^4$, and $S_{\chi \text{PT} }$ is modified as~\cite{Harada:2003jx}
	\begin{equation}\label{hlsf}
		S_{\chi \text{PT} } \to S^{\text{HLS}}_{\chi \text{PT}} 
		= 
		\int_{M^4} 
		\left( \frac{f_P^2}{2} \operatorname{Tr}
		\left( \hat{a}_{\perp \mu} \hat{a}_{\perp}^\mu \right)
		+ \frac{m_V^2}{g^2} \operatorname{Tr}
		\left( \hat{a}_{\parallel \mu} \hat{a}_{\parallel}^\mu \right)
		- \frac{1}{2 g^2}
		\operatorname{Tr}\left( 
		V_{\mu \nu} V^{\mu \nu}
		\right) 
		\right) \,,
	\end{equation}
	where $g$ is the gauge coupling of the HLS. The field strength is defined as $V_{\mu \nu} = \partial_\mu V_\nu - \partial_\nu V_\mu - i [V_\mu, V_\nu]$, and
	\begin{equation}
		V_\mu = \frac{g}{\sqrt{2}} \left( \begin{array}{ccc}
			\frac{1}{\sqrt{2}} (\rho^0 + \omega) & \rho^+ & K^{*+} \\
			\rho^- & \frac{1}{\sqrt{2}} (\omega - \rho^0) & K^{*0} \\
			K^{*-} & \bar{K}^{*0} & \phi
		\end{array} \right)_\mu \,.
	\end{equation}
	In the unitary gauge of the HLS, we choose $h(x)$ such that $\xi_L^\dagger = \xi_R = \xi$, which results in
	\begin{eqnarray}\label{hlsLeading}
		\hat{a}_{\perp \mu} 
		&=& \frac{1}{2i} \left[ 
		\partial_\mu \xi \xi^\dagger - \partial_\mu \xi^\dagger \xi + i \xi {\cal R}_\mu \xi^\dagger - i \xi^\dagger {\cal L}_\mu \xi
		\right] \,, \nonumber \\ 
		\hat{a}_{\parallel \mu} 
		&=& \frac{1}{2i} \left[ 
		\partial_\mu \xi \xi^\dagger + \partial_\mu \xi^\dagger \xi - 2i V_\mu + i \xi {\cal R}_\mu \xi^\dagger + i \xi^\dagger {\cal L}_\mu \xi
		\right] \,,
	\end{eqnarray}
	with $\xi^2 = U$.  In the low-energy limit \( m_V^2 \gg q_V^2 \), where \( q_V \) is the momentum of \( V \), the kinetic terms of \( V_{\mu \nu} \) can be neglected, and the equation of motion for \( V_\mu \) leads to \( \hat{a}_{\parallel} = 0 \). $S ^{\text{HLS}}_{\chi \text{PT}}  $ is then reduced to \( S_{\chi \text{PT}} \).

	Introducing the HLS into $S_{\text{WZW}}$ generates four additional terms, given by~\cite{Fujiwara:1984mp}
	\begin{equation}\label{wzwhls}
		S_{\text{WZW}}^{\text{HLS}} = S_{\text{WZW}} + 
\frac{N_c}{16  \pi^2} 
 \int_{M^4} \sum_{i=1}^4 c_i \operatorname{Tr}\left(
		{\cal I}^{\text{HLS}}_i 
		\right) \,,
	\end{equation}
	where $c_i$ are parameters to be fixed, and
	\begin{eqnarray}
		{\cal I}^{\text{HLS}}_1 &=& i\left( 
		\hat{\alpha}_L^3 \hat{\alpha}_R - \hat{\alpha}_R^3 \hat{\alpha}_L 
		\right) \,, \nonumber \\
		{\cal I}^{\text{HLS}}_2 &=& i \left( \hat{\alpha}_L \hat{\alpha}_R \right)^2 \,, \nonumber \\
		{\cal I}^{\text{HLS}}_3 &=& F_V \left( 
		\hat{\alpha}_L \hat{\alpha}_R - \hat{\alpha}_R \hat{\alpha}_L 
		\right) \,, \nonumber \\
		{\cal I}^{\text{HLS}}_4 &=& \frac{1}{2} \left( 
		\xi^\dagger F_{\cal L} \xi + \xi F_{\cal R} \xi^\dagger 
		\right) \left( 
		\hat{\alpha}_L \hat{\alpha}_R - \hat{\alpha}_R \hat{\alpha}_L
		\right) \,.
	\end{eqnarray}
	The forms are defined by
	\begin{equation}
		\hat{\alpha}_L = \left( \hat{\alpha}_{\parallel \mu} - \hat{\alpha}_{\perp \mu} \right) d x^\mu \,, \quad
		\hat{\alpha}_R = \left( \hat{\alpha}_{\parallel \mu} + \hat{\alpha}_{\perp \mu} \right) d x^\mu \,, \quad 
		F_{\cal V} = d{\cal V} - i {\cal V}^2 \,,
	\end{equation}
	where
	${\cal V} = {\cal V} _\mu dx^\mu$ and 
	 ${\cal V} = V, {\cal L}, {\cal R}$. The equation of motion, neglecting the kinetic terms of $V_{\mu \nu}$, leads to $\hat{a}_ {\parallel} = 0$ and $\hat{\alpha}_R = -\hat{\alpha}_L$, as discussed below eq.~\eqref{hlsLeading}. Hence, $S_{\text{WZW}}^{\text{HLS}}$ is reduced to $S_{\text{WZW}}$ at $q_V^2 = 0$. It is important that 
	 $S_{\text{WZW}}^{\text{HLS}}$ flips sign under the $(-1)^{N_P}$ transformation, while $S _{\chi \text{PT}}^{\text{HLS}}$ remains invariant.
	
	\section{Radiative decays }\label{sec3}
	
In this section, we discuss the radiative decays of $\eta^{(\prime)}$ mesons. The channels $\eta^{(\prime)} \to \gamma \gamma$ have already been thoroughly studied~\cite{Bernstein:2011bx}. Nevertheless, they remain important for fixing the decay constants. To search for further predictions of $S_{\text{WZW}}$, we propose measuring the form factors of $\eta^{(\prime)} \to \pi^+ \pi^- \gamma^{(*)}$.
At $(p_+ + p_-)^2 = 4 m_\pi^2$, with $p_\pm$ being the four-momenta of $\pi^\pm$, the anomalous couplings are uniquely predicted by $S_{\text{WZW}}$, free from vector meson corrections.
Here, $\gamma^*  (V^*)$ stands for the intermediate photon (vector meson).
	
	Due to the finite strange quark mass, the physical $\eta$ and $\eta'$ are mixtures of $\eta_8$ and $\eta_0$. We adopt the single mixing angle scheme of
	\begin{equation}
		\eta = \cos \theta_P \eta_8 - \sin \theta_P \eta_0 \,,~~~
		\eta' = \cos \theta_P \eta_0 + \sin \theta_P \eta_8 \,, 
	\end{equation}
and take  
$	f_\pi = 130~\text{MeV}$, 
	\begin{equation}
		\theta_P = -(25 \pm 2 )^\circ \,,~~
		f_{\eta_8} = (1.59 \pm 0.16) f_\pi \,,~~
		f_{\eta_0} = (1.05 \pm 0.02) f_\pi \,.
	\end{equation}
The above values are obtained by fitting the decays $\eta^{(\prime)} \to \gamma \gamma^{(*)}$ and $\pi^- \pi^+ \gamma^{(*)}$ with the formalism described in the following subsections.
For the HLS parameters, we employ $c_3 = c_4 =1 $ from the VMD, while $c_1 - c_2 = 1$ is taken from $\omega \to \pi^0 \pi^+ \pi^-$~\cite{Harada:2003jx}.
As we will see, this set of parameters indicates that there are no direct three-point vertices that couple $2 \gamma $ and $P$, and the direct four-point vertices for $V\text{-}P\text{-}P\text{-}P$ couplings are forbidden.
	
	\subsection{
		$\eta^{(\prime)} \to \gamma \gamma^{(* )}$ 
	}
\begin{figure}[btbp] 
	\centering
	\includegraphics[width=0.3\linewidth]{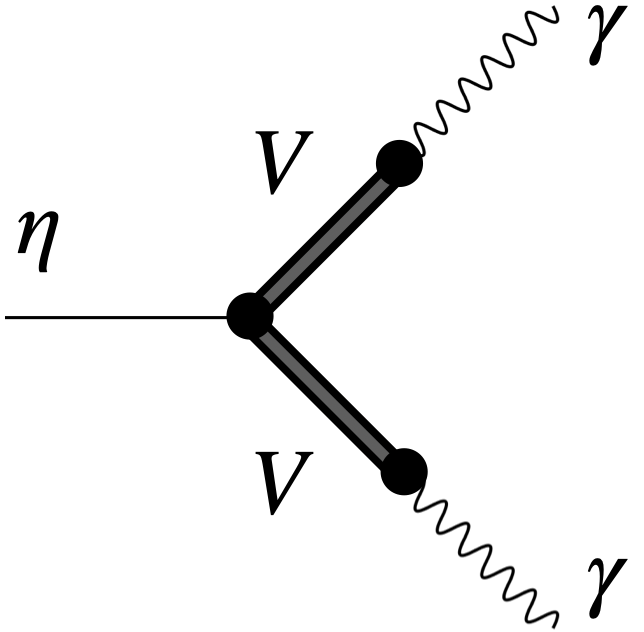} %
	\caption{The  exclusive diagram  for $\eta \to  \pi ^+ \pi^-\gamma$ at the VMD limit with $V = \rho ^ 0 ,\omega , \phi $.  }
	\label{fig:gg}
\end{figure}

	From  $S_{\text{WZW}}^{\text{HLS}}$, the relevant Lagrangian for a meson decay to two photons is given by
	\begin{equation}
		{\cal L}_{PAA} = - \frac{ N_c e^2 }{4 \pi^2 f_{P}}  \left(1 - c_4 \right) \varepsilon^{\mu \nu \alpha \beta} A_{\mu \nu} A_{\alpha \beta} \operatorname{Tr} \left( Q^2 P \right) \,. 
	\end{equation}
	  The vector meson comes into play from the   Lagrangian of 
	\begin{eqnarray}\label{2.4}
		{\cal L}_{PVV} &=& - \frac{N_c}{4 \pi^2 f_{P}} c_3 \varepsilon^{\mu \nu \alpha \beta} \operatorname{Tr}\left( \partial_\mu V_\nu \partial_\alpha V_\beta P \right)\,, \nonumber \\
		{\cal L}_{PVA} &=&  \frac{N_ce }{8 \pi^2  f_{P }} (c_3 - c_4) \varepsilon^{\mu \nu \lambda \sigma} \operatorname{Tr}\left( \{ \partial_\mu V_\nu, Q \partial_\lambda A_\sigma \} P \right) \,.
	\end{eqnarray}
	This is followed by the transition of
	\begin{equation}
		{\cal L}_{VA} = -2\frac{m_V^2}{g^2} e A^\mu \operatorname{Tr} \left( V_\mu Q \right) \,.
	\end{equation}
With $c_3=c_4=1 $, we see that there is no direct coupling of $\eta ^{(\prime )}$ to photons in 3-point vertices. Instead,   photons can only be produced via  $V^*$ as shown in Figure.~\ref{fig:gg}, which explains the name of VMD.

	The decay widths to leading order are then given by
	\begin{equation}
		\Gamma( \eta ^{(\prime)} \to \gamma \gamma ) = \frac{C_{\eta ^{(\prime)}}^2 \alpha_{\text{em}}^2 m_{\eta ^{(\prime)}} ^3}{32 \pi^3} \,.
	\end{equation}
with 	the  anomalous couplings of  
	\begin{eqnarray}
		C_{\eta} &=& \cos \theta_P \frac{1}{\sqrt{3} f_{\eta_8}} - \sin \theta_P \frac{2 \sqrt{2}}{\sqrt{3} f_{\eta_0}}
		= 7.58 \pm 0.18   ~\text{GeV} ^{-1}   \,,\nonumber \\ 
		C_{\eta'} &=& \cos \theta_P \frac{2 \sqrt{2}}{\sqrt{3} f_{\eta_0}} + \sin \theta_P \frac{1}{\sqrt{3} f_{\eta_8}}  = 9.64 \pm 0.08 ~\text{GeV} ^{-1} \,.
	\end{eqnarray}
The numerical values in the right    are extracted from the data and in turn fix the decay constant and the mixing angle. 
	We note that $m_V$ and $g$ are canceled in 
	$C_{ \eta^{(\prime)}} $ 
	 in order to eliminate the $c_i$ dependence at the low energy. The numerical results are summarized in Table~\ref{tab:full}. For comparison, we also listed the values of $\pi^0 \to  \gamma \gamma$. 
 Since that the vector mesons here have $q_V^2 = 0$,  the equation of motion of $V_\mu$  guarantees that $S_{\text{WZW}} ^{\text{HLS}}
 =S_{\text{WZW}}  
 $. The results   with and without the vector meson corrections are essentially the same.

	\begin{table}[htbp]
		\centering
		\begin{tabular}{cccc}
			\hline
			\textbf{Channel} & $\Gamma_{\text{data}}$  
			& $\Gamma_{\text{WZW}}$  & 
$\Gamma_{\text{total}}$  \\
			\hline
	$10^3 \Gamma(  \pi ^0 \to \gamma \gamma)  $    &   		$ 7.71 \pm 0.12 $& $7.78   $& $7.78  $		\\
			$10\Gamma(   \eta  \to \gamma \gamma) $   &  	$5.2 \pm 0.2 $ & $5.1\pm0.2   $  & $5.1\pm0.2   $	\\
			$\Gamma( \eta '  \to \gamma \gamma)  $    &   $4.34 \pm 0.06$ &  $4.38 \pm 0.08$ 	  & $4.38 \pm 0.08$ 		\\
	\hline 
$10^3 \Gamma(   \eta  \to e^+e^-  \gamma) $   &  	$9.0 \pm 0.6 $ & $8.2 \pm 0.4 $  & $8.5  \pm 0.4  $	\\
$10^4 \Gamma(   \eta  \to \mu ^+\mu ^-  \gamma) $   &  	$4.1 \pm 0.5 $ & $2.8  \pm0.1$  & $4.0 \pm0.2  $	\\
$10^2 \Gamma(   \eta'  \to e^+e^-  \gamma) $   &  	$9.2 \pm 0.5 $ & $7.9 \pm 0.1 $  & $8.8 \pm 0.2   $	\\
$10^2 \Gamma(   \eta ' \to \mu ^+\mu ^-  \gamma) $   &  $2.1\pm 0.5$ &$0.76 \pm 0.01$ &$1.66\pm 0.03$  	\\
\hline 
$ 10^2 \Gamma(\eta \to \pi ^+ \pi ^ -  \gamma ) $    & 
			$ 5.6\pm 0.2    $ &  $2.4 \pm 0.1     $ 
			&$5.9 \pm 0.3 $  \\
			$ \Gamma(\eta '  \to \pi ^+ \pi ^ -  \gamma)  $   & 
			$56 \pm 2   $ 
			&  $3.1 \pm 0.1 $ &$56 \pm 1$ \\
	\hline
\multirow{2}{*}{ 
			$ 10^4 \Gamma(\eta \to \pi ^+ \pi ^ -  e^+e^- ) $   }  & 
$  4.07\pm 0.22  $~\cite{BESIII:2025cky} &  
\multirow{2}{*}{ $1.6 \pm 0.1 $ 
}&
\multirow{2}{*}{ $3.7 \pm 0.2 $  }  \\
&$  3.51\pm 0.14 $~\cite{KLOE:2008arm}
\\ 
			$ 10^8 \Gamma(\eta \to \pi ^+ \pi ^ -  \mu ^+\mu ^- ) $    & 
$<52
$~\cite{BESIII:2025cky}  &  $0.50 \pm 0.02 $  
&$1.03   \pm 0.05          $   \\
 $ 10   \Gamma(\eta ' \to \pi ^+ \pi ^ -  e^+e^- ) $    & 
$ 4.5\pm 0.2   $ &  $0.24\pm 0.01$ 
&$4.1\pm 0.1 $    \\
$ 10^3   \Gamma(\eta ' \to \pi ^+ \pi ^ -  \mu ^+\mu ^- ) $    & 
$  3.6\pm  0.8  $ &  $0.56 \pm 0.01 $ 
&$4.1\pm 0.1   $  
			\\
			\hline
		\end{tabular}
		\footnotetext{from Box anomaly}  
		\caption{
The decay widths of anomalous decays are given in units of keV, where  \( \Gamma_{\text{data}} \) denotes the experimental decay widths, and  \( \Gamma_{\text{WZW}} \) and \( \Gamma_{\text{total}} \) represent the theoretical predictions without and with vector meson corrections, respectively. 
		}
		\label{tab:full}
	\end{table}

It is straightforward to extend the study to $\eta^{(\prime)} \to \ell^+ \ell^- \gamma$ with $\ell  = e $ or $ \mu $. The partial decay width is given by 
\begin{equation}
	\frac{ \partial \Gamma }
	{\partial s _ \ell }  
	= \frac{C_{{\eta ^{(\prime)}}}  ^2 \alpha_{\text{em}}^3 }{96 \pi^4 m _{\eta^{(\prime)}}  ^3 s_\ell }
	|F _{\eta ^{(\prime)}} |^2 
	\left(
	m_{\eta ^{(\prime)}}^2 - s_\ell 
	\right) ^3 \beta_\ell (
	3 - \beta_\ell ^2 
	) \,,
\end{equation}
with $s_\ell = (p_{\ell^+} + p_{\ell^-})^2$ and $\beta_\ell =
\sqrt{1 - 4m_\ell^2/s_\ell}$. The form factors, 
normalized as $F_{\eta ^{(\prime)}}(0) = 1$, read:
\begin{eqnarray}
	F_\eta  (s_\ell)  &=& 
	\frac{1}{6 \sqrt{3}C_\eta }
	\left[
	\left(
	\frac{9}{D_\rho } 
	+\frac{1}{D_ \omega }
	-\frac{4}{D_\phi }
	\right)\frac{\cos\theta_P}{
		f_{\eta_8}  
	}
	-\sqrt{2} \left(
	\frac{9}{D_\rho } 
	+ \frac{1}{ D_\omega}
	+ \frac{2}{D_\phi}
	\right)
	\frac{\sin \theta_P }{ f_{\eta_0} }
	\right]
	\,,\nonumber\\
	F_{\eta'}  (s_\ell)  &=& 
	\frac{1}{6 \sqrt{3}C_{ \eta '} }
	\left[
	\sqrt{2} \left(
	\frac{9}{D_\rho } 
	+ \frac{1}{ D_\omega}
	+ \frac{2}{D_\phi}
	\right)
	\frac{\cos  \theta_P }{ f_{\eta_0} }
	+
	\left(
	\frac{9}{D_\rho } 
	+\frac{1}{D_ \omega }
	-\frac{4}{D_\phi }
	\right)\frac{\sin\theta_P}{
		f_{\eta_8}  
	}
	\right]
	\,,
\end{eqnarray}
where $D_V^{-1} = m_V^2 / ( m_V^2 - i m_V \Gamma_V(s_\ell) - s_\ell )$ for $V = \rho^0, \omega, \phi$, and~\cite{Zhang:2012gt}
\begin{equation}
	\Gamma_V(s_i) 
	=\left(\frac{\Gamma_{V} \sqrt{s_i}}{m_V}\right)\left(\frac{1-\frac{4 m_i^2}{s_i }}{1-\frac{4 m_i^2}{m_V^2}}\right)^{\frac{3}{2}} \Theta\left(s _i - 4 m_i^2\right)
\end{equation}
with $\Gamma_V$ the total decay width of $V$ and $ i =  \ell, \pi$.  
The numerical results are summarized in Table~\ref{tab:full}, and good agreements have been found with the experiments.

The experiments have measured the slope of the form factors, found to be 
$( \Lambda , \Lambda' ) = ( 0.716 \pm 0.011, 0.79 \pm 0.04)$ 
in units of GeV,  
defined by 
$\partial F_{ \eta^{(\prime)}}  /\partial s_\ell = 1/ ( \Lambda ^{(\prime)} )^2$ at $s_\ell = 0$~\cite{ParticleDataGroup:2024cfk}.  
In the experiments, it is fitted from a single dipole behavior of $|F_{\eta^{(\prime)}}|^2 \propto  ( 1 - s_\ell / \Lambda^{(\prime)2}  ) ^{-2}$.  
In the VMD, we have that  
$( \Lambda, \Lambda')=(  0.76 \pm 0.01 , 0.82 \pm 0.01 )$~GeV,  
as expected from $\Lambda^{(\prime)} \approx m_V$.  
The deviation from the experimental value is within $5\%$ and may arise from higher-order loop corrections~\cite{Hanhart:2013vba}.  
We stress that the predictions of $C_\eta$ and $C_{\eta'}$ would not be affected by the uncertainties in the HLS parameters.

	\subsection{
		$\eta ^{(\prime )}    \to   \pi ^+ \pi ^-  \gamma^{(*)}  $ 
	}

	\begin{figure}[btbp] 
		\centering
		\includegraphics[width=0.3\linewidth]{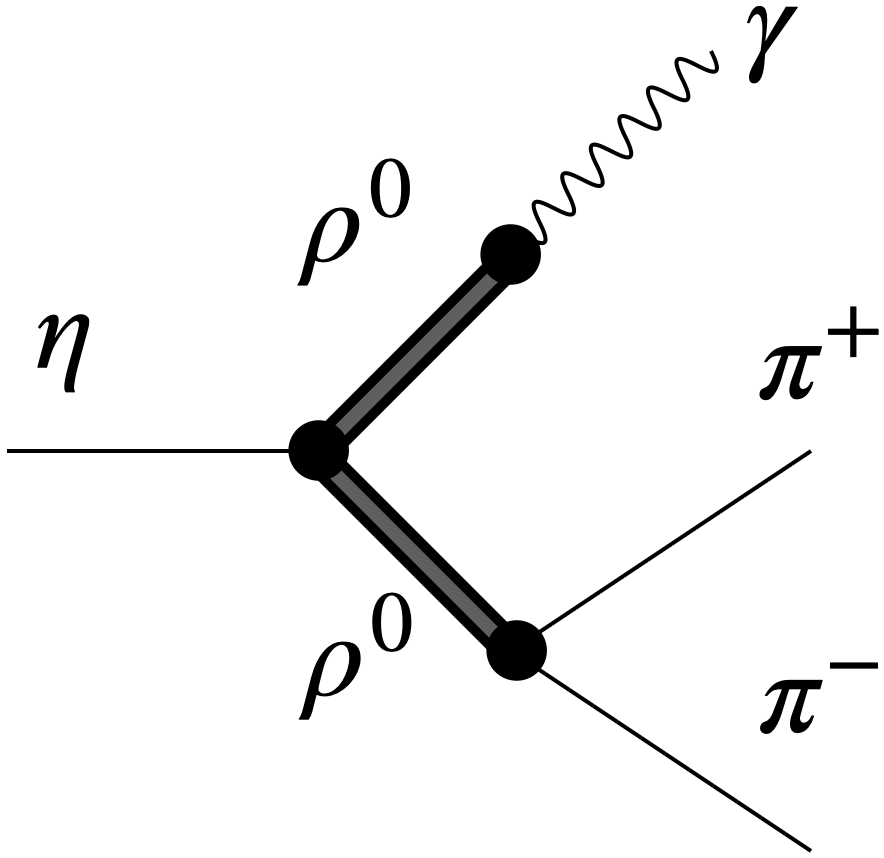} %
		\includegraphics[width=0.3\linewidth]{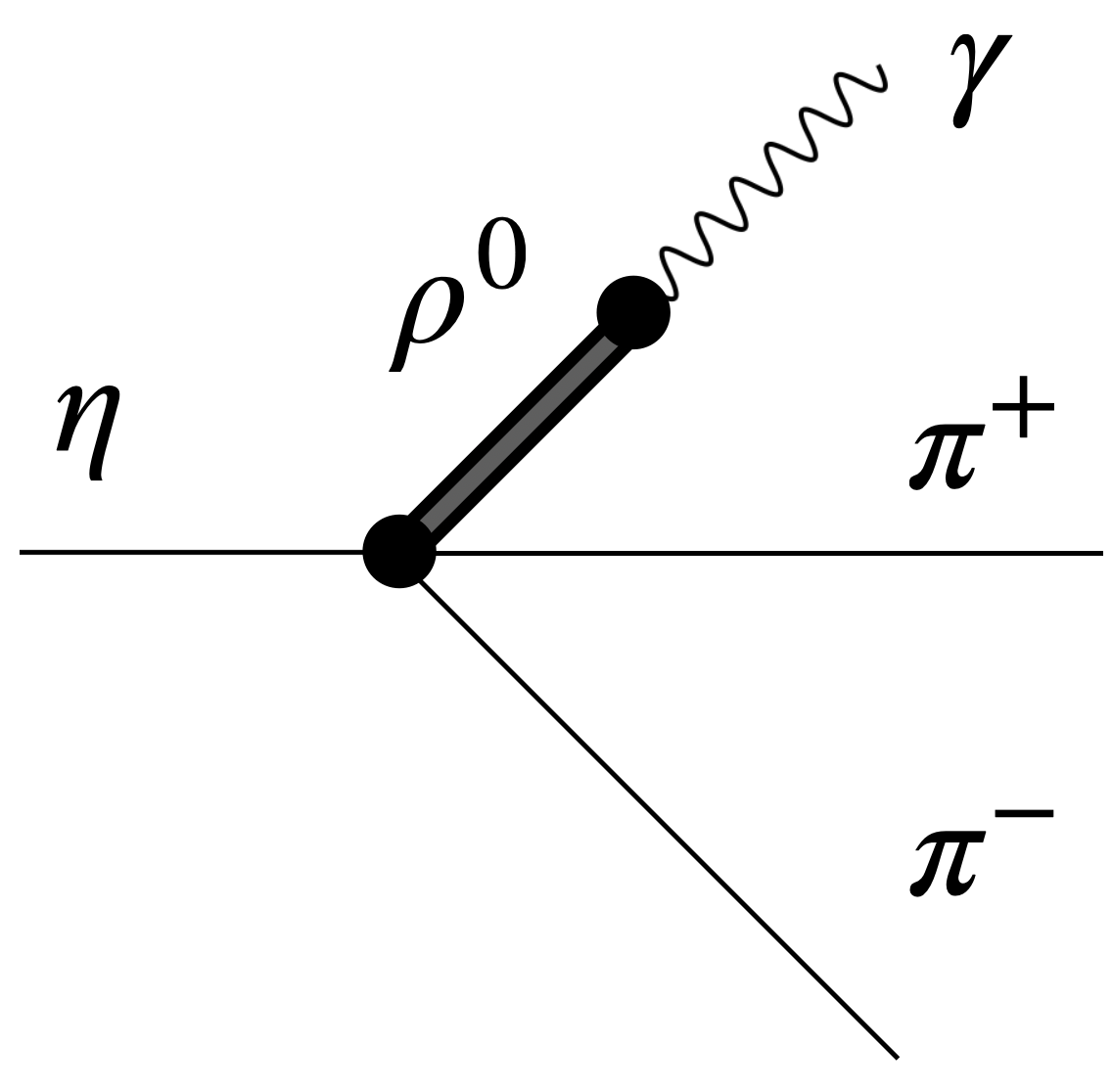}
	\includegraphics[width=0.3\linewidth]{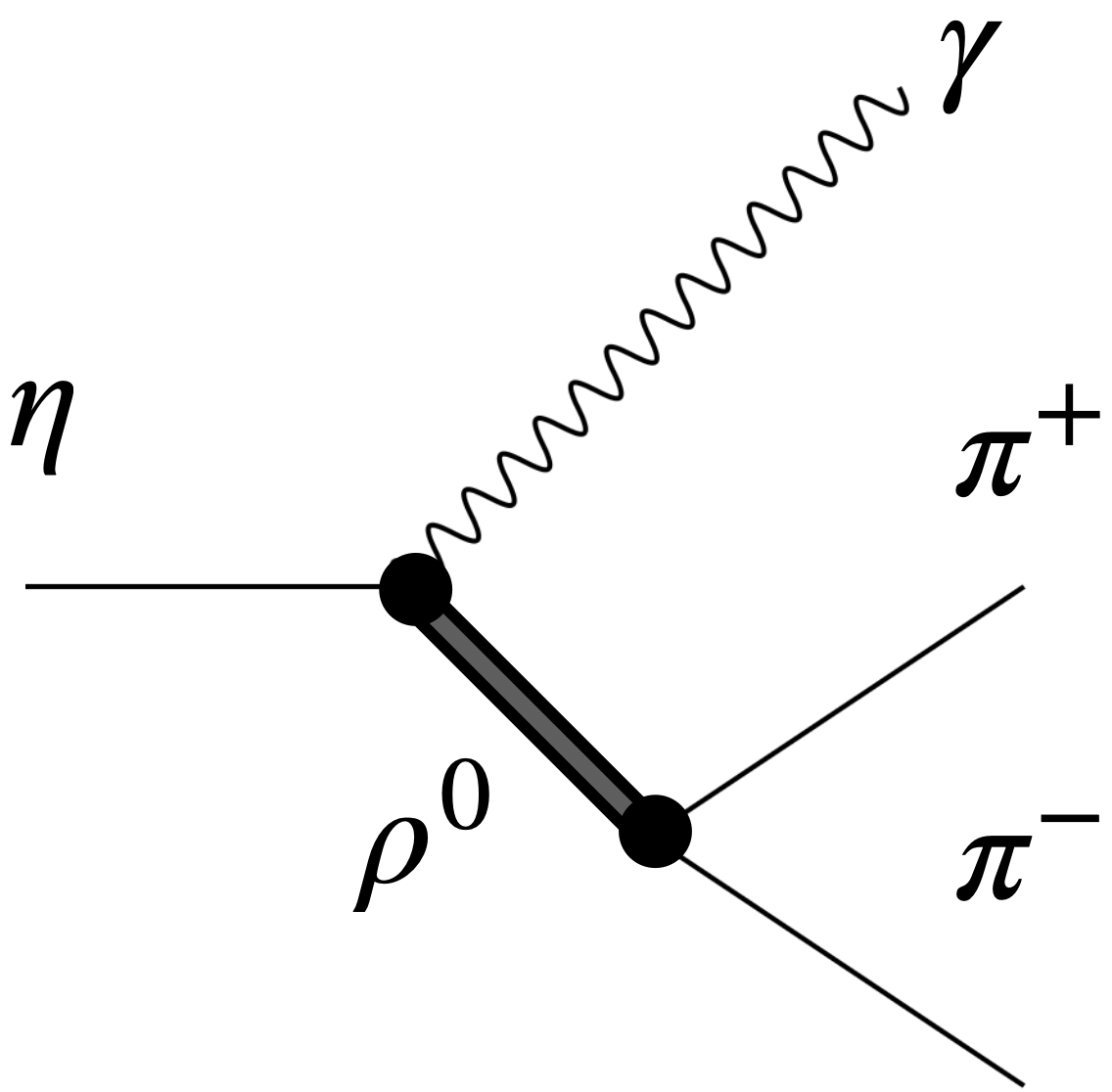}   
		\caption{Resonance diagrams for $\eta \to  \pi ^+ \pi^-\gamma$.}
		\label{fig:HLS}
	\end{figure}

To further examine the anomalous sector, we turn to 
$\eta^{(\prime)} \to \pi ^+ \pi ^- \gamma^{(*)}$, which are induced by   box-diagrams  at the quark level~\cite{Picciotto:1993aa}. 
	Expanding $S_{\text{WZW}}^{\text{HLS}}$, the Lagrangian responsible for 
	the direct transition of 
	$\eta \to \pi^+ \pi^- \gamma$ is given by
	\begin{equation}
		{\cal L}_{PPPA} = - \frac{i N_c}{3 \pi^2 f_P^3} \left( 1 - \frac{3}{4} (c_1 - c_2 + c_4) \right) \varepsilon^{\mu \nu \alpha \beta} A_\mu \operatorname{Tr} \left( Q \partial_\nu P \partial_\alpha P \partial_\beta P \right) \,.
	\end{equation}
The decays can also be induced by the resonant diagrams depicted  in Figure~\ref{fig:HLS}. 
  The $\eta^{(\prime)} \to V^*V^* $ vertex  is  found in eq.~\eqref{2.4}, while the one for $\eta ^{(\prime)}  \to V^*  \pi^+ \pi^-$ is governed by
	\begin{equation}\label{4po}
		{\cal L}_{VPPP} = -i \frac{N_c}{4 \pi^2 f_P^3} \left( c_1 - c_2 - c_3 \right) \varepsilon^{\mu \nu \alpha \beta} \operatorname{Tr} \left( V_\mu \partial_\nu P \partial_\alpha P \partial_\beta P \right) \,,
	\end{equation}
and $c_1 -c_2 = c_3$ leads to $	{\cal L}_{VPPP} = 0 $. 
	The transitions of $V ^* \to \pi^+ \pi^-$ are described by
	\begin{equation}
		{\cal L}_{VPP} = i \frac{m_V^2}{g^2 f_P^2} \operatorname{Tr} \left( V ^\mu [ \partial_\mu P , P ] \right) \,.
	\end{equation} 
There is one additional type of diagrams that comes from $\eta \to \pi ^\pm  \rho^{\mp} ( \to  \pi^\mp  \gamma) $ with $\rho^{\mp }$ as the intermediate meson. Nevertheless, $\eta \to \pi ^\pm \rho ^\mp $ are  forbidden by the G-parity and can therefore be discarded.

	We label the momenta as $\eta \to \pi^+(p_+) \pi^-(p_-) \gamma(q, \epsilon)$. Piecing together all the contributions results in 
	\begin{equation}
\mathcal{A}_{\eta^{(\prime)}} = i e C_{\eta^{(\prime)}}^{[\pi\pi]} 
F_V ^{[\pi\pi]}
\varepsilon_{\mu \nu \rho \lambda} p_+^\mu p_-^\nu q^\rho \epsilon^\lambda \,,
	\end{equation}
	and the anomalous couplings of 
	\begin{eqnarray}\label{212}
		C_\eta^{[\pi\pi]} &=& \left( \frac{\cos \theta_P / f_8 - \sqrt{2} \sin \theta_P / f_0}{  \sqrt{6} \pi^2 f_\pi^2} \right) 
		 = 21.4 \pm 0.5 ~\text{GeV}^{-3}  
		 \,, \nonumber\\
		C_{\eta'}^{[\pi\pi]} & =& \left( \frac{\sqrt{2} \cos \theta_P / f_0 + \sin \theta_P / f_8}{  \sqrt{6 } \pi^2 f_\pi^2} \right) 
		= 17.9 \pm 0.3 ~\text{GeV} ^{-3}
		 \,.
	\end{eqnarray}
Because $\pi^+ \pi^-$ is antisymmetric and forms an isospin triplet, only  $\rho^0$ contributes to the resonant diagrams.
Hence,  \( \eta \) and \( \eta' \) share the same form factor \( F_V^{[\pi\pi]} \), given by 
	\begin{equation}
		F_V^{[\pi\pi]} (s_\pi ) = 1 - \frac{3}{4} \left( c_3 + c_4\right) \left(   \frac{s_\pi }{s_\pi  - \overline{m}_\rho^2 } 
		+ \delta  \frac{s_\pi }{s_\pi - \overline{m} _\omega^2 } \right) 
		\,,
	\end{equation}
with $\overline{m}_V ^2 = m_\rho ^2 - i \Gamma _V(s_i)  m _V$.  Here, we have included  $\delta$ to  take  account the isospin breaking, which may   arise from the $\rho^0-\omega$ mixing. 
 The partial decay widths are 
	\begin{equation}\label{par}
		\frac{d \Gamma}{d s_\pi } = \frac{\alpha_{\text{em}}}{192 \pi^2} E_\gamma^3 
\beta_\pi ^3 
		s_\pi  \left|C_{\eta^{(\prime)}}^{[\pi\pi]}\right|^2
		\left|
		F_V^{[\pi\pi]} 
		\right|^2 
		 \,,
	\end{equation}
	where  $E_\gamma = ( m^2_{\eta^{(\prime)}} - s_\pi )/ ( 2 m_{\eta^{(\prime)}} ) $ is the photon energy and 
	$\beta_{ \pi } = \sqrt{
		1 - 4 m_{\pi } ^2 /s _ \pi  
	}	$. The numerical results of the decay widths, compared to the data, are listed in Table~\ref{tab:full}, and good agreements have been found.

\begin{figure}[bt] 
		\centering
		\includegraphics[width=0.6\linewidth]{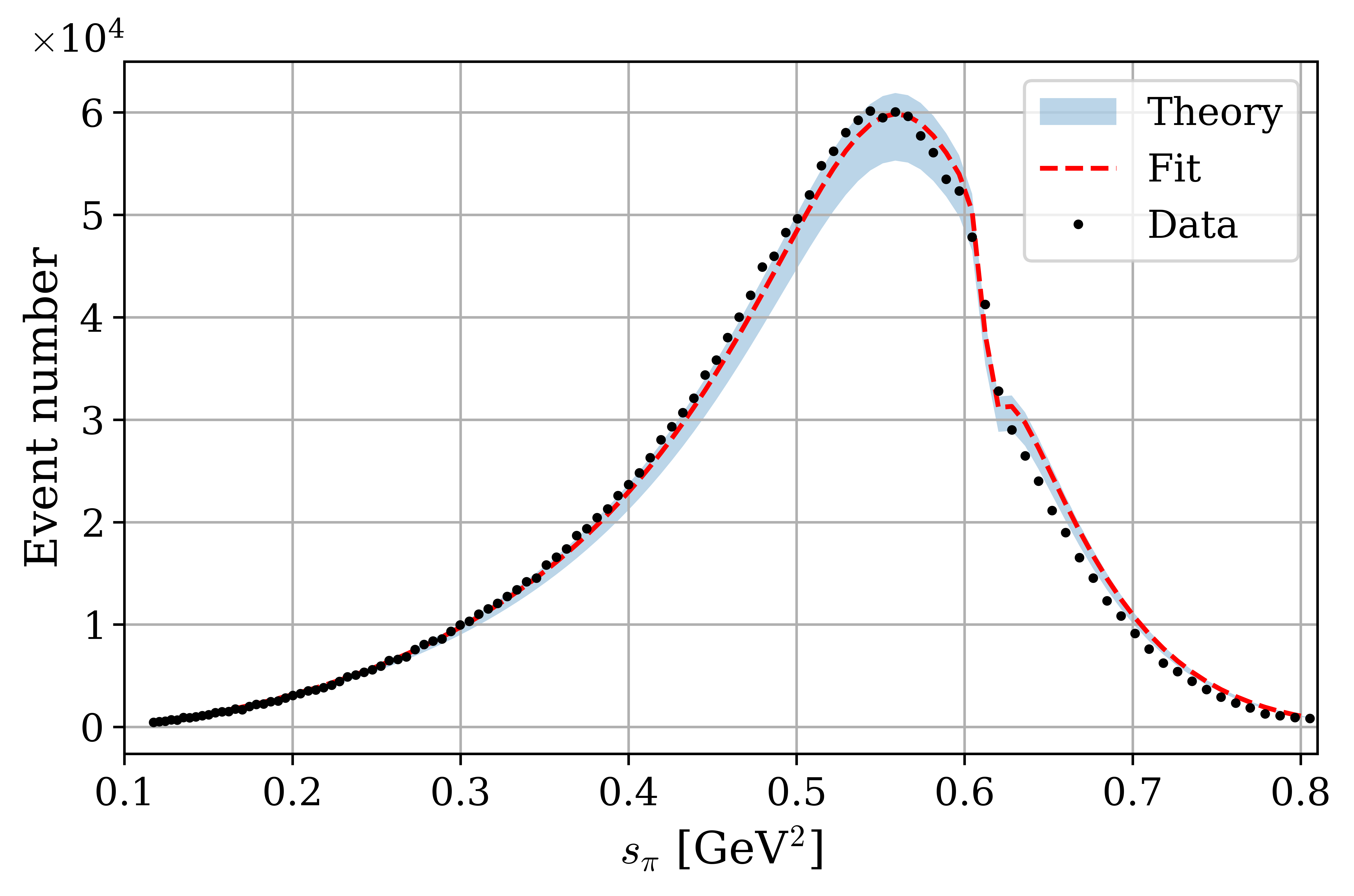} %
		\caption{Extraction of $C_{\eta’}^{[\pi\pi]}(\mathrm{data})$ from Ref.~\cite{BESIII:2015zpz}, where the reconstructed partial decay width multiplied by the cross section and luminosity is presented as the red line with $C_{\eta’}^{[\pi\pi]}(\mathrm{data}) = (18.2 \pm 0.1)$~GeV$^{-3}$,  and  the light blue band is from  the parameter input of eq.~\eqref{212}.
			}
		\label{fig:fit}
	\end{figure}

It is interesting to point out that the experiment at BESIII has examined the partial decay width and obtained the branching ratio~\cite{BESIII:2017kyd}
\begin{equation}
	{\cal B}   _{\text{box}}^{\text{exp}}(\eta' \to \pi^+ \pi^- \gamma)
	= (0.245 \pm 0.021)    \%  \,,
\end{equation}
where the subscript “box" indicates   the box anomaly contributions. However, this value is smaller by an order of magnitude compared to our result: 
\begin{equation}\label{br}
	{\cal B}_{\text{box}}^{\text{theory}}(\eta' \to \pi^+ \pi^- \gamma)
	= (1.65 \pm 0.07)  \%  \,.
\end{equation}
Using the provided experimental data points~\cite{BESIII:2017kyd},  
we perform a $\chi^2$ fit with eq.~\eqref{par} by treating $C_{\eta'}^{[\pi\pi]}$ as an unknown while setting $c_3 = c_4 = 1$.  
From the extraction, we found 
\begin{equation}
	C_{\eta'}^{[\pi\pi]} (\mathrm{data}) = (18.2 \pm 0.1)~\text{GeV}^{-3} \,,
\end{equation}
and the reconstructed ${\cal B}_{\mathrm{box}}^{\text{exp}}$ to be $(1.70 \pm 0.05)\%$,  
in good accordance with eqs.~\eqref{212} and \eqref{br}.  
The overall fit is presented in Figure~\ref{fig:fit}.  
A bump around $s_\pi \approx m_\omega^2$ is caused by isospin-breaking effects from intermediate $\omega$, with  
(Re$(\delta)$, Im$(\delta)$)  
=  
$(-7.5 \pm 0.9,\, 7.4 \pm 0.8) \times 10^{-3}$,  
which is consistent with the size of the  $\rho^0$–$\omega$ mixing~\cite{Chen:2017jcw,Mitchell:1996dn}.

\begin{figure} [bt]
		\centering
		\includegraphics[width=0.45 \linewidth]{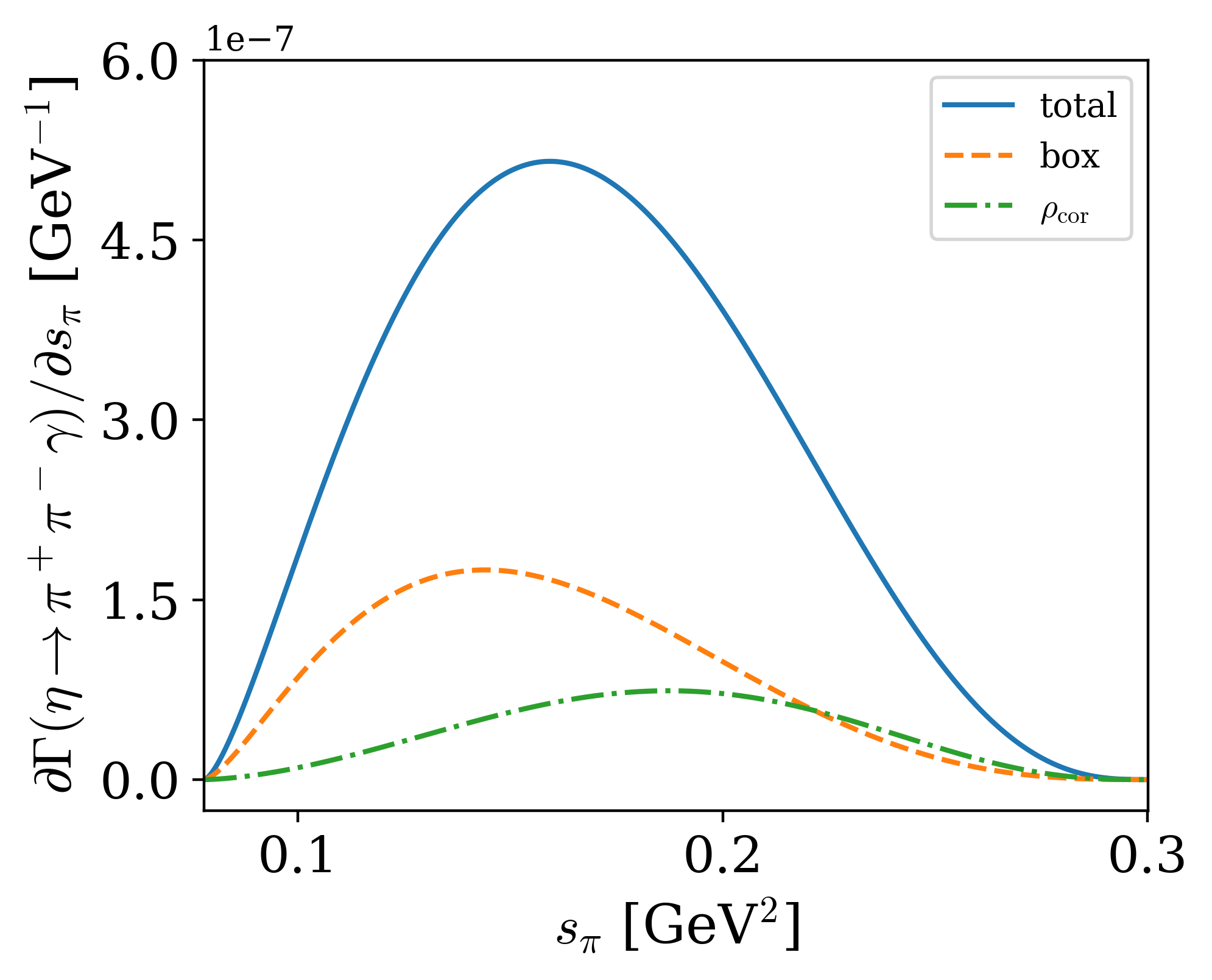}
		\includegraphics[width=0.45 \linewidth]{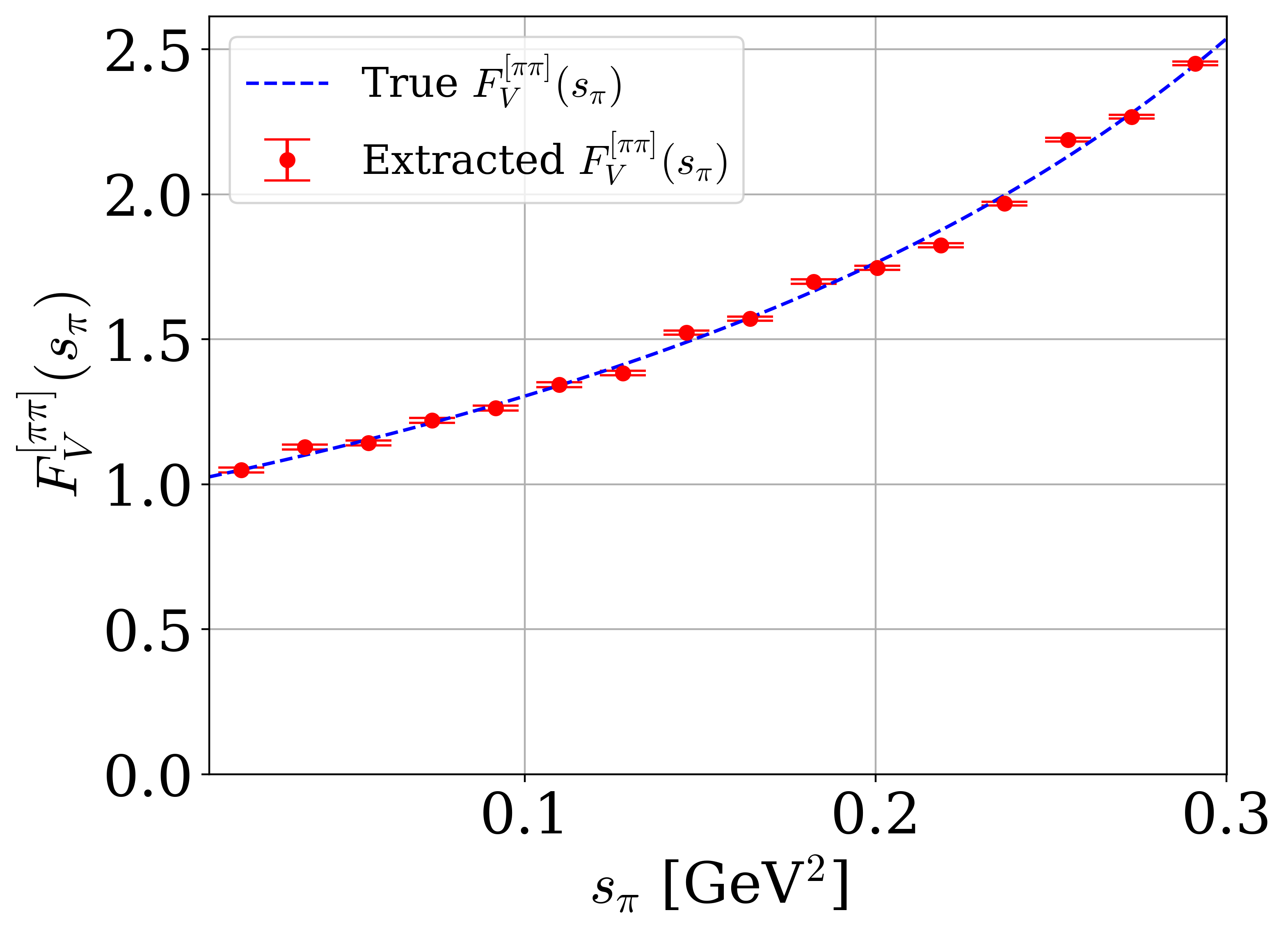}
		\caption{The left hand side is the theoretical partial decay width of $\eta \to \pi ^+ \pi ^- \gamma$ while the right is a simulation on extractions of $F_V^{ [\pi\pi]} (s)$.}
		\label{fig:your_label}
	\end{figure}

	In contrast to $\eta' \to \pi^+ \pi^- \gamma$,  the box anomaly plays the leading role in $\eta \to \pi^+ \pi^- \gamma$. In Figure~\ref{fig:your_label}, we   depict the partial decay width of $\eta \to \pi^+ \pi^- \gamma$. In the figure, the line labeled “total" represents the total decay width by adding up  the box anomaly and $\rho^0$ correction, while the lines labeled “box" and “$\rho_{\text{cor}}$" correspond to  the partial decay width exclusively coming from each, respectively. The two of them interfere constructively, and   the partial decay width is dominated by the former. Physically, this is due to the fact that $m_\eta$ is much smaller than $m_{\eta'}$, and hence $S_{\text{WZW}}$ leads to a much better estimation.
	
The branching fraction of \( \mathcal{B}^{\text{exp}}(\eta \to \pi^+ \pi^- \gamma) = (4.28 \pm 0.07)\% \) has been measured with high precision at BESIII via the decay chain \( J/\psi \to \eta \gamma \)~\cite{BESIII:2021fos}. However, the extraction of the box anomaly contribution has not yet been performed.  
A simulated fit of \( F_V^{[\pi\pi]}(s) \) is shown on the right-hand side of Figure~\ref{fig:your_label}, assuming a total of \( 8 \times 10^4 \) events and a systematic uncertainty of 2\%~\cite{BESIII:2021fos}. The pseudo-data points are generated with \(   c_3 + c_4 = 2  \).  
By extracting the values of \( C_{\eta}^{[\pi\pi]} \) and \( c_3 + c_4 \) from the simulated data, we find that the precision of \( C_{\eta}^{[\pi\pi]} \) reaches 1\%, while the precision of \( c_3 + c_4 \) reaches 4\%.  
We emphasize that \( C_{\eta}^{[\pi\pi]} \) is independent of vector meson corrections, making it an excellent observable for cleanly testing the anomalous coupling.  
On the other hand, at future colliders such as the  STCF, the number of produced $J/\psi$ mesons is estimated to be a hundred times higher than at BESIII, and the expected number of observed events may reach \( 10^7 \)~\cite{Achasov:2023gey}, improving the precision of \( C_{\eta}^{[\pi\pi]} \) to \( 10^{-3} \).

To probe deeper the  form factor, it may be useful to consider the off-shell process for the photon  $\eta ^{(\prime)} \to \pi ^ + \pi ^ - \ell   ^+\ell  ^-$, where the amplitude is given by 
\begin{equation}
\mathcal{A}_{\eta^{( \prime)}} = i e^2 C_{\eta^{( \prime)}}^{[\pi\pi]} 
F_V ^{[\pi\pi]} 
\varepsilon_{\mu \nu \rho \lambda} p_+^\mu p_-^\nu q^\rho  
\frac{1}{s_l}
\overline u_e   \gamma_\mu v_e  \,, 
\end{equation}
and $C_{\eta^{( \prime)}}^{[\pi\pi]} $ is given by eq.~\eqref{212}. On the other hand,  $F_V^{[\pi\pi]} $  acquires  dependence on $s_\ell =  ( p_{\ell ^+} + p_{\ell ^-} )^2  $: 
\begin{eqnarray}\label{fvee}
	F_V^{[\pi\pi]}  &=& 1 - \frac{3(c_1- c_2  ) }{4} 
	\frac{s_\ell }{s_\ell   -\overline{m}_\rho^2  }
	-\frac{3}{4} c_4 \frac{s_\pi}{s_\pi - \overline{m}_\rho^2}
+\frac{3}{4} c_3  
 \frac{m_\rho^2 (s_\pi +s_\ell )}{(   s_\pi - \overline{m}_\rho^2 )(  s_\ell - \overline{m}_\rho^2 )}
	\,. 
\end{eqnarray}
We note that $c_4 = c_3$ has been used in the literature. 
We   integrate out   degrees of freedom in angles and arrive at 
\begin{eqnarray}
\frac{	
	\partial ^2\Gamma 
}{
	\partial  s_\pi  \partial s_\ell 
}
&=&
\left| C_{0^{(\prime)}}^{[\pi\pi]}  F_V^{[\pi\pi]} 
\right|^2 
\frac{ \alpha_{\text{em}}^2  
\beta_\pi^3 \beta _\ell 
\sqrt{\lambda^3}
}{
512  \pi ^3 m_{\eta^{(\prime)}}^3 
}
\frac{s_\pi}{9s_\ell^2 }  
\left(
s_\ell   + 2 m_\ell ^2 
\right)
\end{eqnarray}
	where 
 $\lambda =
m_{\eta^{(\prime)}} ^4 + s_\pi  ^2 + s^2 _\ell -2 m_{\eta^{(\prime)}}^2 s_\pi  - 2 m_{\eta^{(\prime)}}^2 s_\ell - 2 s _\pi s_\ell  
$.
The predicted  decay widths 
with $c_1 -c_2 = c_3 =c_4 = 1$ 
are listed in Table~\ref{tab:full}.

\begin{figure}[bt] 
	\centering
	\includegraphics[width=0.4\linewidth]{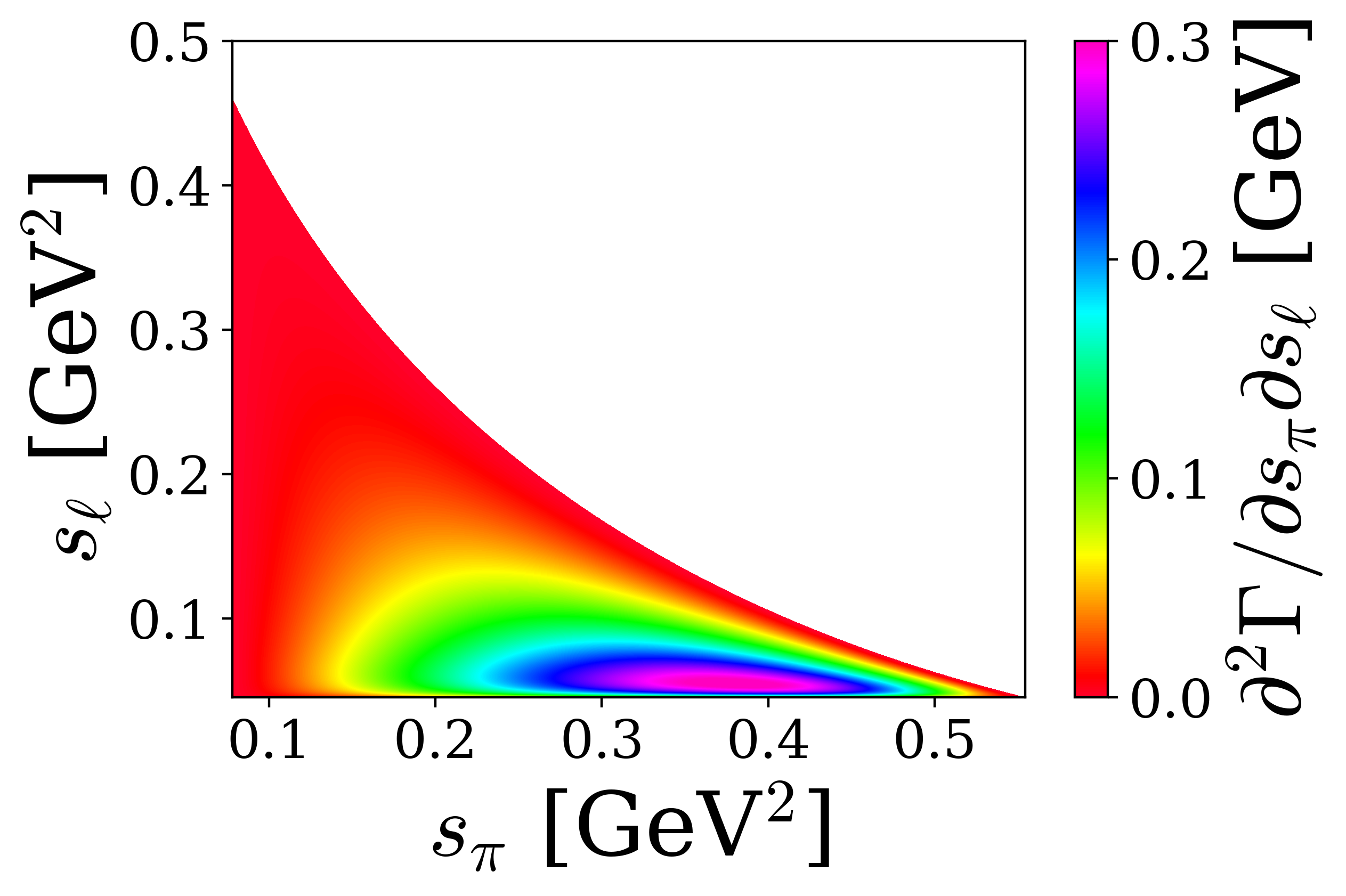} %
	\includegraphics[width=0.4\linewidth]{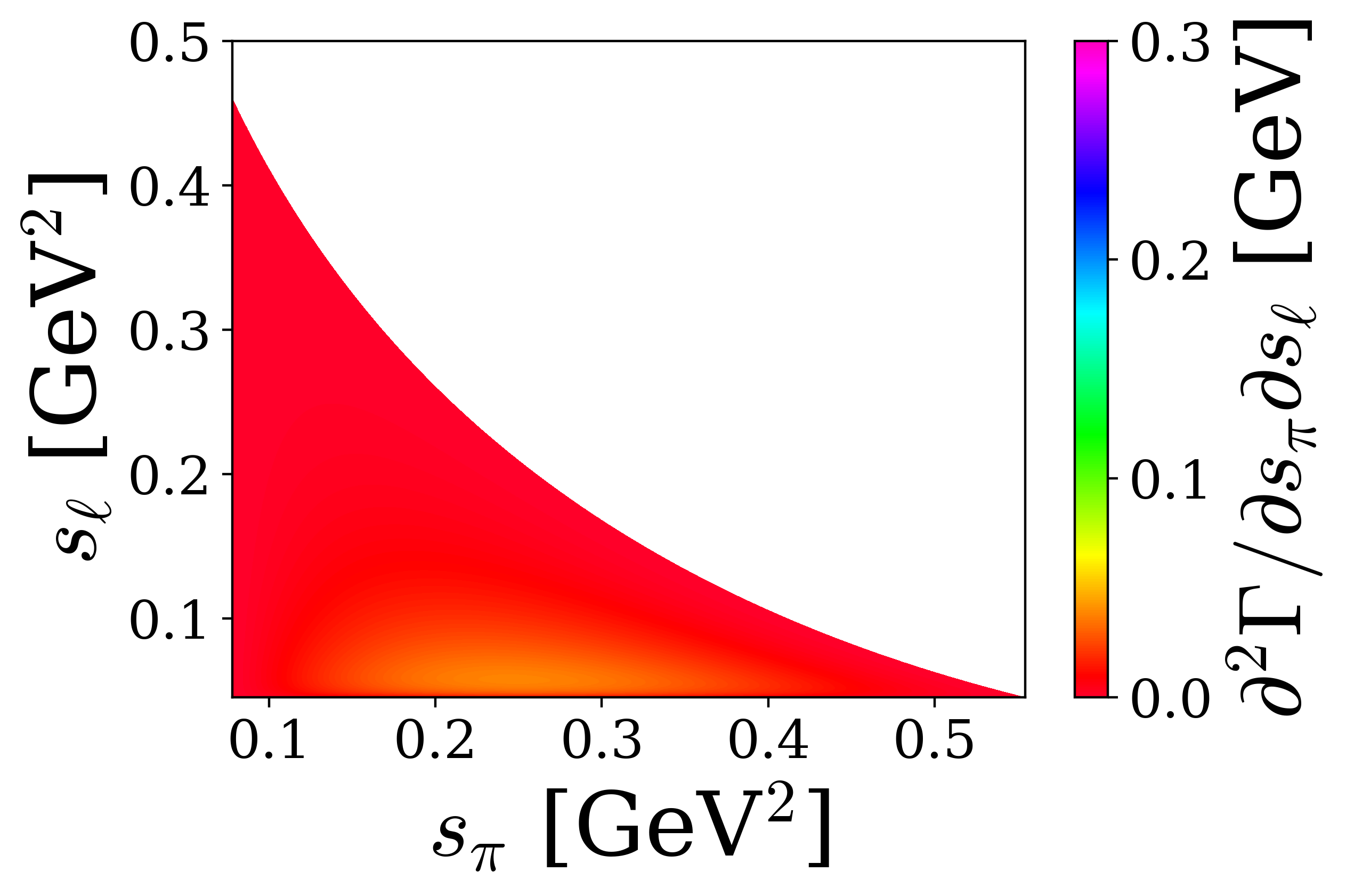} %
	\caption{
		Partial decay width for $\eta' \to \pi^+ \pi ^- \mu  ^+ \mu ^-$, where the left~(right) figure corresponds to  the case with~(without) vector  meson corrections.}
	\label{fig:C11}
\end{figure}

Focusing on extracting $c_1 - c_2$ in eq.~\eqref{fvee}, processes involving $e^+ e^-$ are not very useful, as the partial decay width is proportional to $s_\ell^{-1}$. In the high $s_\ell$ region, the partial width becomes very small, rendering the $s_\ell$-dependent structure in $F_V^{[\pi\pi]}$ difficult to resolve. On the other hand, the branching fraction of $\eta \to \pi^+ \pi^- \mu^+ \mu^-$ is too small to be observed. We conclude that the most suitable channel to probe $c_1 - c_2$ is $\eta' \to \pi^+ \pi^- \mu^+ \mu^-$, for which the Dalitz plot is shown in Figure~\ref{fig:C11}.
In the figure, we display the cases with $c_1 - c_2 = c_3 = c_4 = 1$ and $c_1 - c_2 = c_3 = c_4 = 0$ on the left and right, respectively. The large discrepancy between the two plots highlights the important role of vector meson corrections. The BESIII collaboration found $(c_1 - c_2) = 0.01 \pm 0.045$ and $c_3 = 0.98 \pm 0.40$, which is in $2\sigma$ tension with the theoretical value $c_1 - c_2 = 1$~\cite{BESIII:2024awu}.
Interestingly,
by setting $c_1 -c_2 = c_3 = 1 $, 
 BESIII also reported that the vector meson masses responsible for producing  different leptons and pions differ. In particular, the effective masses for $V^* \to e^+ e^-$ and $V^* \to \mu^+ \mu^-$ are $954.3 \pm 95.0$~MeV and $649.4 \pm 66.3$~MeV, respectively~\cite{BESIII:2024awu}. In contrast, isospin symmetry suggests that only $\rho^0$ mesons contribute to $\eta^{(\prime)} \to \pi^+ \pi^- \gamma^{(*)}$, with $m_V = m_\rho$.
The raw data is not yet publicly available, and a future experimental revisit is strongly recommended.

Before ending this section, we note that the anomalous sector can also be examined with two photons in the final states, such as $\eta^{(\prime)} \to \pi^+ \pi^- \gamma \gamma$~\cite{Knochlein:1996ah}. In particular, $\Gamma_{\text{WZW}}$ are  calculated  to be $6 \times 10^{-6}$~keV and $1.2 \times 10^{-6}$ keV for $\eta \to \pi^+ \pi^- \gamma \gamma$ and $\eta' \to \pi^+ \pi^- \gamma \gamma$, respectively.  
However, the momentum dependence of the form factors is much more complicated, and extractions of the anomalous couplings are not feasible at the current stage.

	\section{
		Kaon semileptonic decays 
	} \label{sec4}
	In this section, we discuss the semileptonic decay of kaons with two pions in the final states. Although these processes violate parity and are   allowed by $S_{\chi \text{PT} }$, only terms in $S_{\text{WZW}}$ break the symmetry of the coordinate transformation of  $\vec{x} \to -\vec{x}$, and unique behaviors can be identified~\cite{Ametller:1993hg}. The amplitude of \( {K}^{+} \to \pi^+  \pi^- \ell ^+  \nu_\ell   \) is given by~\cite{NA482:2010dug}
	\begin{eqnarray}
		{\cal A}({K}^{+} \to \pi^+ \pi^- e^+ \nu_e) 
		&=& \frac{G_F}{\sqrt{2}} V_{us}^* \overline{u}_\nu \gamma^\mu (1- \gamma_5) v_e 
		\Bigg[ \frac{2H^+}{m_K^3} \epsilon_{\mu \nu \alpha \beta} p_K^\nu p_+^\alpha p_-^\beta \\
		&&+ \frac{i}{m_K} \left( F^+ (p_+ + p_-)_\mu + G^+ \left( p_{  +} - p_{ -} \right)_\lambda + R^+ \left( p_{ {e}} + p_\nu  \right)_\lambda \right) \Bigg] \,, \nonumber
	\end{eqnarray}
	where \( F, G, H, R \) are the form factors. 
The decay of \( K^0 \to \pi^- \pi^0 e^+ \nu_e \) is parameterized in a similar manner with $F^+,G^+,H^+,R^+$ replaced by $F^0,G^0,H^0,R^0$. 
	The values of \( F, G,  R \) can be derived  from \( S _{\chi \text{PT}} \), while \( H  \) is exclusively contributed by \( S_{\text{WZW}} \). It can be seen from the fact that the term proportional to \( H  \) flips sign under the coordinate transformation \( \vec{x} \to -\vec{x} \) due to the presence of the totally antisymmetric tensor.
Due to the parity conservation, $H $ must come from the vector part of the  current while the rest from the axial part.
We conclude the parity-conserving amplitude is included in, and only in, $H$.
	
	Explicitly, the governing Lagrangian that contributes to \( H^{0,+} \) is given by
	\begin{eqnarray}
		{\cal L}_{PPPW} 
		&=& \frac{i N_c g_2}{6\sqrt{2}\pi^2 f_P^3}
\left[
1 - \frac{3}{4}
\left(
c_1 - c_2 + c_4 
\right)
\right]
		\epsilon^{\mu \nu \alpha \beta}
		W_\beta^+ \operatorname{Tr}
		\left( 
		\partial_\mu P \partial_\nu P \partial_\alpha P 
		 T_+  
		\right)\,,\nonumber\\
	{\cal L}_{VW} &=& -\frac{1}{\sqrt{2}} \frac{m_V^2}{g^2} g_w  W^+_ \mu \operatorname{Tr} \left( V^\mu T_+  \right) \,,\nonumber\\
	{\cal L}_{PVW} &=&   \frac{N_cg_2}{16 \sqrt{2} \pi f_P} (c_3- c_4) \epsilon^{\mu \nu \lambda \sigma} \operatorname{Tr}\left( \{ \partial_\mu V_\nu, T_+ \partial_\lambda W^+ _\sigma \} P \right) 
	\end{eqnarray}
The ratios between the couplings   are exactly the same with photons as only the vector part of the gauge boson contributes 
as discussed in the previous paragraph.

\begin{figure}[bt] 
	\centering
	\includegraphics[width=0.2\linewidth]{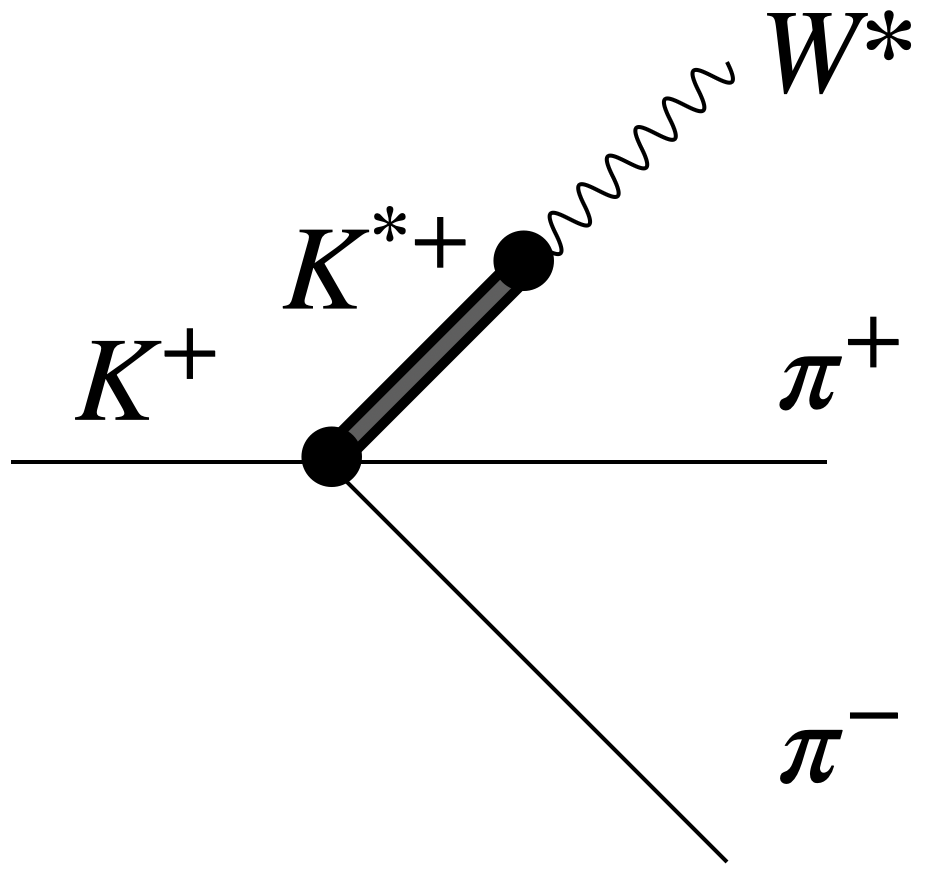} ~
	\includegraphics[width=0.2\linewidth]{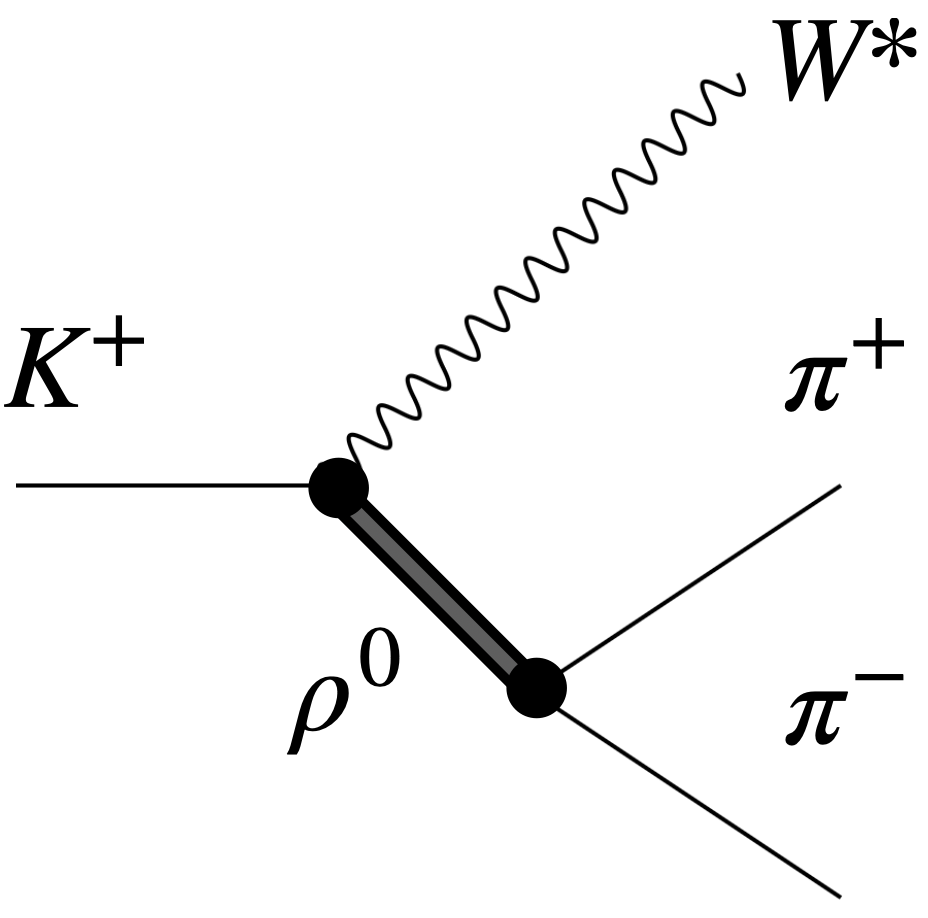} ~  
	\includegraphics[width=0.2\linewidth]{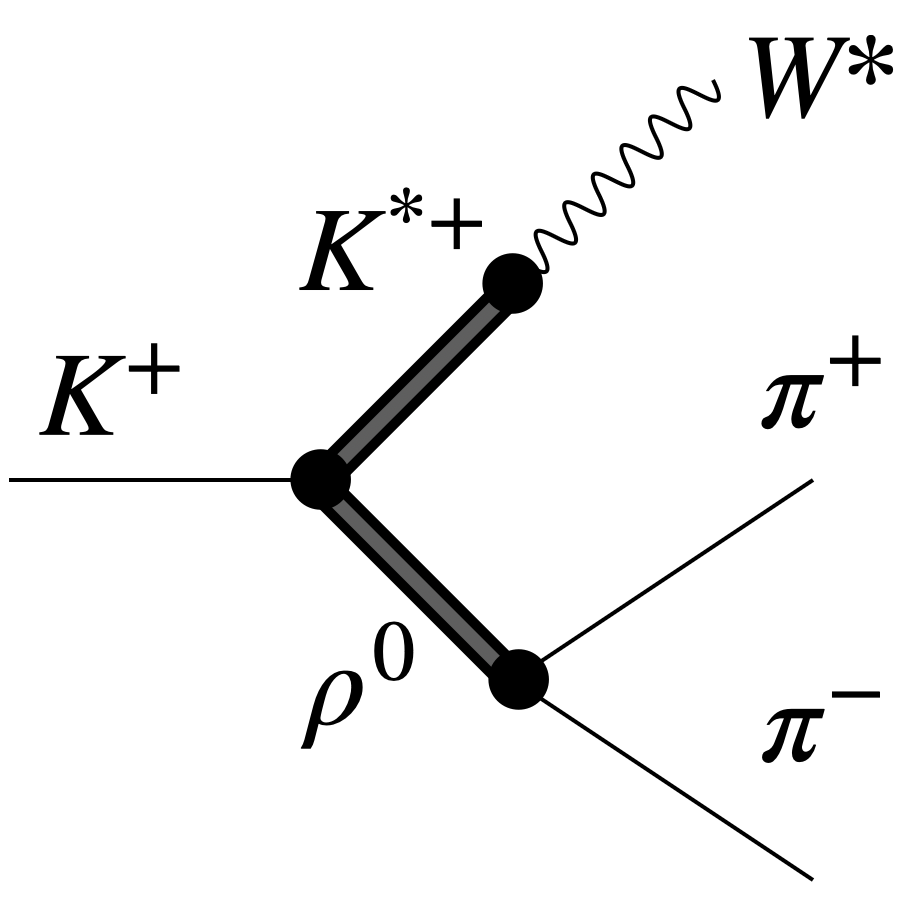}   
	\includegraphics[width=0.25\linewidth]{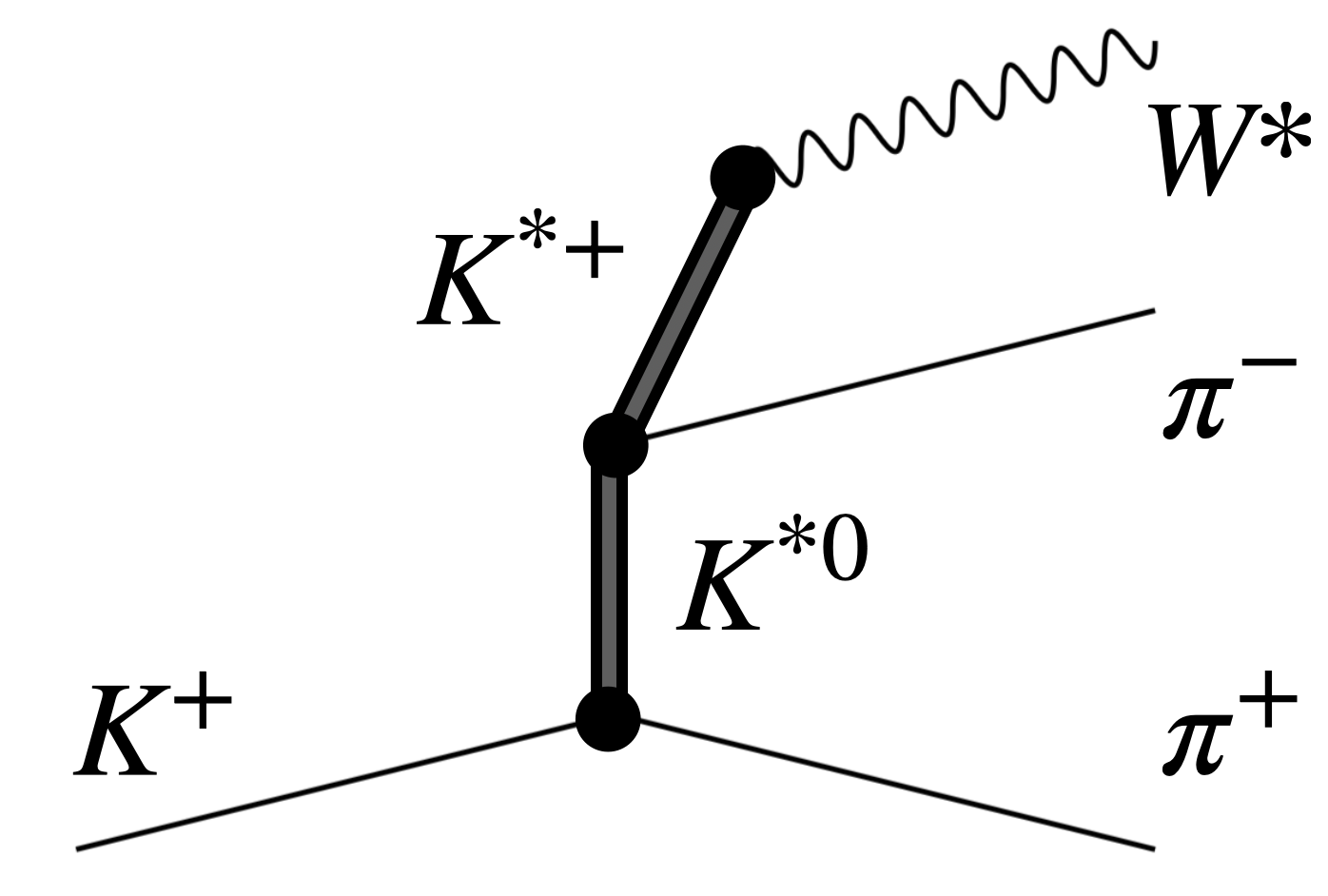}   
	\caption{Diagrams for $K^+ \to \pi^+ \pi ^ - W^* $.}
	\label{fig:KKK}
\end{figure}

\begin{figure}[bt] 
	\centering
	\includegraphics[width=0.4 \linewidth]{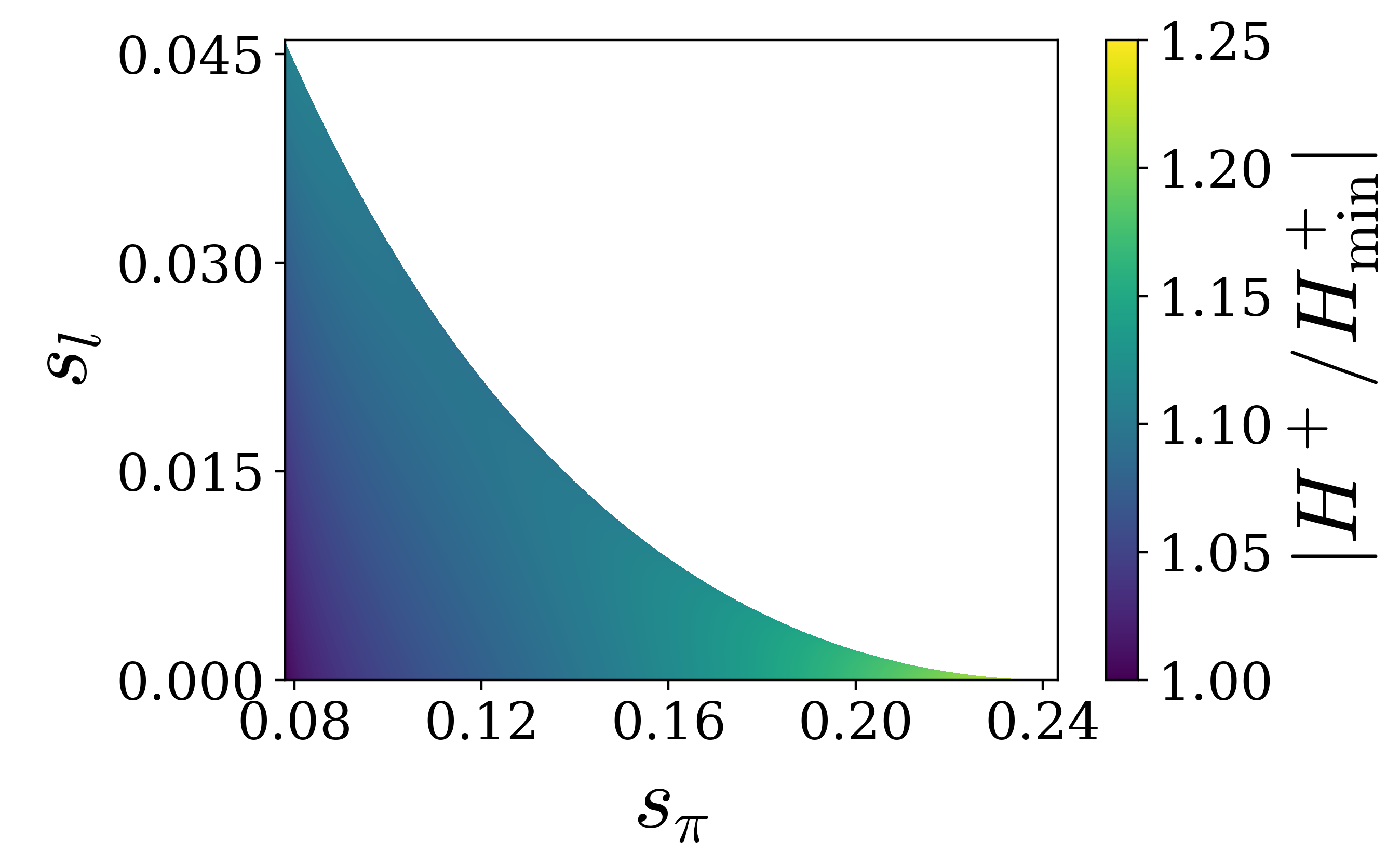} %
	\caption{Vector meson corrections of  to the form factors of $|H^+|$ with $c_1 -c_2 =c_3 =c_4=1$, where   $H_{\min}^+$ is defined as  $H_{\min}^+= H(0,0,-m_\pi^2/2 )$. }
	\label{fig:K_form}
\end{figure}

The considered Feynman diagrams are depicted in Figure~\ref{fig:KKK}. The fourth diagram induces a topology that does not occurs in $\eta \to \pi ^+ \pi ^- \gamma^*$. 
Setting $c_3 =c_4$, 
we arrive at  
\begin{eqnarray}\label{eqre}
H^+ 
(s_\pi , s_\ell ,q\cdot p_-)
&= & H^+_{\text{min}}
\Bigg\{ 
1 - \frac{3}{4}\left(
c_1 -c_2 -c_3 
\right)
\frac{ s_\ell }{s_\ell  - \overline{m} _{K^*}^2  }
 \\
&&-\frac{3}{4}
c_3 
\left[ 
2 -
\frac{
m_{K^*}^2
}{\overline{m} _{K^*}^2 - s_\ell } 
\left(
 \frac{m_\rho^2}{\overline{m}_\rho^2 - s_\pi }
 + 
  \frac{
  {m} _{K^*}^2
  }{\overline{m} _{K^*}^2 -  (q+p_-)^2}
\right)
\right]
\Bigg\},\nonumber\\
H^0
(s_\pi , s_\ell ,q\cdot p_-)
&=& \frac{\sqrt{2}}{2}
\left(
H^+ 
( s_\pi , s_\ell ,q\cdot p_-)+ 
H^+ 
\left(s_\pi , s_\ell ,
\frac{1}{2} \left( m_K^2 -s _\ell -s_\pi - 2 
q\cdot p_-\right) 
\right) \right) \,, \nonumber 
\end{eqnarray}
where $q  = p_\ell  + p_\nu $ and $H^+_{\text{min}}= H^+(0,0,-m_\pi^2/ 2 )$ corresponds to the anomalous coupling   from $S_{\text{WZW}}$. 
The last term 
in $H^+$ 
induces an  angle dependence by the substitution of 
\begin{equation}
q\cdot p  _ - = \frac{1}{4}
\left(
m_K^2 - s_\pi -s_\ell  - \beta_\pi \sqrt{\lambda} \cos\theta _\pi  
\right)\,,
\end{equation}
where $\theta_\pi$ is defined as the angle between  $(\vec{p}_+ + \vec{p _-})$ 
defined at the kaon rest frame and $\vec{p}_+$ at the center-of-mass frame of $\pi^+\pi^-$.
From the equation, the maximal correction from the vector mesons can be as large as 
$25\%$ depicted in Figure~\ref{fig:K_form}.

	Using the input \( f_K = 156 \)~MeV and taking the low-energy limit of  $s_\ell = s_\pi = 0$, we find that~\cite{Bijnens:1994ie}
	\begin{equation}
( H^+_{\text{min}})_ {\text{theory}}
  = 
\frac{
m_K^3
}{2\pi^2 f_\pi^2 f_K }		=
		-2.31\,.
	\end{equation}
It is remarkably close to the experimental result~\cite{NA48:2004oba,NA482:2012cho}
	\begin{equation}
		(H^+)_{\text{exp}} = -2.27 \pm 0.10\,,
	\end{equation}
where the experiments have  assumed that  $(H^+)_{\text{exp}}$  is independent of the phase space.
	Here, the sign is fixed by taking \( G^+ \) to be positive. Using eq.~\eqref{eqre}, we find
	\begin{equation}
( H^0_{\text{min}})_ {\text{theory}}= \sqrt{2} ( H^+_{\text{min}})_ {\text{theory}}  = -3.27\,.
	\end{equation}
	On the other hand, the data gives \( | H^0 / G^0 | = -0.32 \pm 0.14 \). With  the isospin relation \( G^0 = \sqrt{2} G^+ \), we find that
	\begin{equation}
		(H^0)_{\text{exp}} = - 2.2 \pm 1.0\,,
	\end{equation}
	which is consistent with $H^0$ from $S_{\text{WZW}}$.

As stressed, the $H^+$ originates exclusively from the vector coupling. As a result, it does not interfere with the other form factors in the total decay width due to the  parity conservation. We can, therefore, compute the branching fraction contributed by $H^+$ exclusively.
The partial decay width is given by
\begin{equation}
	\frac{\partial^3 \Gamma }{
		\partial s_\pi \partial s_\ell \partial \cos \theta_\pi
	}
	\left|
	H
	\right|^2
	\frac{G_F^2 |V_{us}|^2}{
		2 ^{14} \pi ^5 m_K ^9
	}
	\sqrt{\lambda_K^3} \beta_\pi ^3
	s_\ell s_\pi
	\frac{(1 -z_\ell)^2(
		2 +  z_\ell )  }{3}
	\left(
	1 - \cos ^2 \theta_\pi
	\right)
	\,.
\end{equation}
Here, $z_\ell = m_\ell ^2 / s_\ell $ and $\lambda_K = m_K^4+ s_\ell ^2 + s_\pi^2 - 2 m_K^2 s_\ell - 2 m_K^2 s_\pi - 2 s_\pi s_\ell$.
By taking $c_1 - c_2 = c_3 = c_4=1$, we find the contributions of $S_{\text{WZW}}^{\text{HLS}} $ to the branching fractions as
\begin{eqnarray}
	{\cal B}_{\text{PC}} ( K^+ \to \pi^+ \pi ^- e^+ \nu_e ,
	\pi^+ \pi ^- \mu ^+ \nu_ \mu
	)&=&
	\left( 1.8\times 10 ^{-8}, 1.2 \times 10 ^{-9}
	\right) \,,
	\nonumber\\
	{\cal B}_{\text{PC}} ( K_L^0 \to \pi^\pm  \pi ^0  e^\mp  \nu ,
	\pi^\pm  \pi ^0  \mu ^\mp  \nu
	)&=&
	(1.3 \times 10 ^{-7},
	9.1 \times 10 ^{-9}
	) \,.
\end{eqnarray}
The subscripts denote that only the parity-conserving part of the branching fractions are considered.
On the other hand, if we turn off the vector meson corrections, the branching fractions become
$(1.1\times 10 ^{-8}, 7.6 \times 10 ^{-10})$ and
$(8.9 \times 10 ^{-8},6.3\times 10 ^{-9})$, respectively.
We see that the vector meson corrections are around 50\% and cannot be neglected.

 Comparing with the experimental values:
\begin{eqnarray}
	{\cal B}_{\text{exp}}( K^+ \to \pi^+ \pi ^- e^+ \nu_e ,
	\pi^+ \pi ^- \mu ^+ \nu_ \mu
	)&=&
	\left( ( 4.247\pm 0.024)\times 10 ^{-5}, ( 1.4\pm 0.9) \times 10 ^{-5}
	\right) 
\,,
	\nonumber\\
	{\cal B}_{\text{exp}}( K_L^0 \to \pi^\pm  \pi ^0  e^\mp   \nu
	)&=&( 5.20\pm0.11) \times 10 ^{-5} 	  \,,
\end{eqnarray}
we conclude that the anomalous couplings  have little impact on the total branching fractions at the current precision. The best way to probe the WZW term is by employing the full decay distributions and extracting its interference with the leading form factors~\cite{NA482:2012cho}. In the future, it will be important to apply a phase-space dependence when extracting the form factors, as we have found sizable dependencies on $s_\pi$ and $s_\ell$.
	
Before ending this section, we note the anomalous couplings can also induce the double neutrino decays of  
$K^- \to \pi^- \pi^0 \nu \overline{\nu}$ via intermediate $Z$ bosons. However, to extract the corresponding anomalous couplings and form factors, it is necessary to reconstruct all the final-state 4-momenta, which is not feasible at colliders in the near future.

\section{Conclusion}
\label{sec5}

We have investigated the identification of the  WZW  Lagrangian at colliders such as BESIII and the STCF with  the  HLS  framework. 
We have examined the anomalous decays $\eta^{(\prime)} \to \pi^+\pi^-\gamma^{(*)}$, focusing on the extractions of anomalous couplings. The theoretical predictions yield anomalous couplings 
$C_{\eta}^{[\pi\pi]} = 21.4 \pm 0.5~\text{GeV}^{-3}$ and 
$C_{\eta'}^{[\pi\pi]} = 17.9 \pm 0.3~\text{GeV}^{-3}$.
Experimental measurements at BESIII gave a branching ratio 
$\mathcal{B}_{\text{box}}^{\text{exp}}(\eta' \to \pi^+\pi^-\gamma) = (0.245 \pm 0.021)\%$, 
which is significantly lower than our theoretical result, 
$\mathcal{B}_{\text{box}}^{\text{theory}}(\eta' \to \pi^+\pi^-\gamma) = (1.65 \pm 0.07)\%$. 
However, using the experimental data to perform a $\chi^2$ fit, we have  obtained 
$C_{\eta'}^{[\pi\pi]}(\text{data}) = 18.2 \pm 0.1~\text{GeV}^{-3}$, 
and the reconstructed branching ratio becomes 
$\mathcal{B}_{\text{box}}^{\text{exp}}(\eta' \to \pi^+\pi^-\gamma) = (1.70 \pm 0.05)\%$, 
which
 is an order of magnitude larger  than the previous experimental determination and 
 aligns well with the theoretical value.
For the decay $\eta \to \pi^+\pi^-\gamma$, simulation studies suggest that the precision of $C_{\eta}^{[\pi\pi]}$ can reach 1\%, and that of $c_3 + c_4$ can reach 4\%. 
At future colliders such as the STCF, with an expected $10^7$ events, the precision of $C_{\eta}^{[\pi\pi]}$ may be improved to the level of $10^{-3}$.
The BESIII collaboration found vector meson correction parameters 
$(c_1 - c_2) = 0.01 \pm 0.045$ and $c_3 = 0.98 \pm 0.40$, 
which show a $2\sigma$ tension with the assumption $c_1 - c_2 = c_3 =1 $. 
Additionally, by setting $c_1 - c_2 = c_3 = c_4$ 
in $\eta \to \pi^+ \pi^- e^+ e^-$, 
the BESIII collaboration  reported different vector meson masses responsible for 
$V^* \to e^+ e^-$: $954.3 \pm 95.0~\text{MeV}$, and 
$V^* \to \mu^+ \mu^-$: $649.4 \pm 66.3~\text{MeV}$. 
In contrast, within our framework, only $\rho^0$ contributes to this process in the isospin symmetry limit and  $m_V = m_\rho$, 
suggesting that further theoretical and experimental investigations are needed.

In ${K}^{+} \to \pi^+ \pi^- \ell^+ \nu_\ell$, the anomalous couplings contribute uniquely to the form factor $H$, which can be extracted from parity-conserving decay distributions. Although predictions at $s_\pi = s_\ell = 0$  closely matched experimental values (e.g., $H^+ = -2.31$ versus $-2.27 \pm 0.10$), we have identified substantial corrections arising from intermediate vector meson states, reaching magnitudes of up to 25\%. The assumption of a constant $H^+$ made in the experimental extraction may not be valid, and we recommend revisiting these experiments to refine the determination of the form factor.

\acknowledgments

We would like to thank Shuang-Shi  Fang, Hai-Bo Li  and 
Ben-Hou Xiang 
for the useful discussions about the BESIII data points.
This work was supported in part by 
the National Key Research and Development Program of China under Grant No. 2020YFC2201501;
the National Science Foundation of China (NSFC) under Grants No. 1821505, 12205063, and 12347103; and the Strategic Priority Research Program and Special Fund of the Chinese Academy of Sciences.

\end{document}